\documentclass[10pt,journal]{IEEEtran} 
\makeatletter
\usepackage{subfigure}
\usepackage{times}
\usepackage{amsthm}
\newcommand{\subparagraph}{}
\theoremstyle{definition}

\long\def\@makecaption#1#2{\ifx\@captype\@IEEEtablestring%
\footnotesize\begin{center}{\normalfont\footnotesize #1}\\
{\normalfont\footnotesize\scshape #2}\end{center}%
\@IEEEtablecaptionsepspace
\else
\@IEEEfigurecaptionsepspace
\setbox\@tempboxa\hbox{\normalfont\footnotesize {#1.}~~ #2}%
\ifdim \wd\@tempboxa >\hsize%
\setbox\@tempboxa\hbox{\normalfont\footnotesize {#1.}~~ }%
\parbox[t]{\hsize}{\normalfont\footnotesize \noindent\unhbox\@tempboxa#2}%
\else
\hbox to\hsize{\normalfont\footnotesize\hfil\box\@tempboxa\hfil}\fi\fi}
\makeatother
\usepackage{booktabs}
\usepackage{array}
\newcolumntype{P}[1]{>{\centering\arraybackslash}p{#1}}
\usepackage{lscape}
\usepackage{comment}
\usepackage{verbatim}
\usepackage{ltablex}
\usepackage[numbers]{natbib}
\usepackage{lscape}
\usepackage[ruled]{algorithm2e}
\usepackage{wrapfig}
\usepackage{color,soul}
\usepackage{longtable}
\usepackage{amssymb,amsmath}
\usepackage{multirow}
\usepackage{varwidth}
\usepackage{colortbl}
\usepackage[usenames,dvipsnames]{xcolor}
\usepackage[pdftex]{graphicx}
\usepackage{stfloats}
\usepackage{graphicx}
\usepackage{tikz}
\usetikzlibrary{er}
\usetikzlibrary{shapes,snakes}
\usetikzlibrary{shapes.gates.logic.US,trees,positioning,arrows}
\usepackage{pgfplots}
\pgfplotsset{width=18cm, height=6cm}
\usepackage{pifont}

\newcommand{\xmark}{\text{\ding{55}}}
\usepackage[normalem]{ulem}

\usepackage[switch,columnwise]{lineno}
\usepackage{pgf-pie}
\SetAlFnt{\small}
\SetAlCapFnt{\small}
\SetAlCapNameFnt{\small}
\SetAlCapHSkip{0pt}
\IncMargin{-\parindent}
\newcolumntype{P}[1]{>{\centering\arraybackslash}p{#1}}
\tikzstyle{chart}=
[legend label/.style={font={\scriptsize},anchor=west,align=left},
legend box/.style={rectangle, draw, minimum size=5pt},
axis/.style={black,thin,->},
axis label/.style={anchor=east,font={\tiny}}]

\tikzstyle{bar chart}=[
chart, bar width/.code={
    \pgfmathparse{##1/2}
    \global\let\bar@w\pgfmathresult},
bar/.style={thin, draw=black},
bar label/.style={font={\bf\small},anchor=north},
bar value/.style={font={\footnotesize}},
bar width=.75,]

\tikzstyle{pie chart}=
[chart,
slice/.style={line cap=round, line join=round, thin, draw=black},
pie title/.style={font={}},
slice type/.style 2 args={
    ##1/.style={fill=##2},
    values of ##1/.style={}}]

\pgfdeclarelayer{background}
\pgfdeclarelayer{foreground}
\pgfsetlayers{background,main,foreground}

\newcommand{\tempsum}{0}

\newcommand{\pye}[3][]{
    \begin{scope}[#1]
        \pgfmathsetmacro{\curA}{90}
        \pgfmathsetmacro{\r}{1}
        \def\c{(0,0)}
        \def\tempsum{0}
        \foreach \v/\s in{#3}{
          \pgfmathparse{\v+\tempsum}
          \global\let\tempsum=\pgfmathresult}
        \foreach \v/\s in{#3}{
            \pgfmathsetmacro{\deltaA}{\v/\tempsum*360}
            \pgfmathsetmacro{\nextA}{\curA + \deltaA}
            \pgfmathsetmacro{\midA}{(\curA+\nextA)/2}
            \path[slice,\s] \c
            -- +(\curA:\r)
            arc (\curA:\nextA:\r)
            -- cycle;
            \pgfmathsetmacro{\d}{1.2}
            \begin{pgfonlayer}{foreground}
                \path \c -- node[pos=\d,pie values,values of \s]{$\v$} +  (\midA:\r);
            \end{pgfonlayer}
            \global\let\curA\nextA}
    \end{scope}}

\newcommand{\legend}[2][]{
    \begin{scope}[#1]
        \path
        \foreach \n/\s in {#2}
        {++(0,-5pt) node[\s,legend box] {} +(2pt,0) node[legend label]     {\n}};
    \end{scope}}
\usepackage{pgfplots}
\definecolor{ac1}{HTML}{b88b4d}

\usepackage{hyperref}
\usepackage{enumitem}
\setlist[itemize]{leftmargin=*}

\newcommand{\cb}{\cellcolor{black!20}}

\begin{document}
\pagenumbering{arabic}


\title{Diversity-By-Design for Dependable and Secure Cyber-Physical Systems: A Survey}
\author{Qisheng Zhang, Abdullah Zubair Mohammed, Zelin Wan, Jin-Hee Cho, \IEEEmembership{Senior Member, IEEE}, and Terrence J. Moore, \IEEEmembership{Member, IEEE}\IEEEcompsocitemizethanks{\IEEEcompsocthanksitem Qisheng Zhang, Zelin Wan, and Jin-Hee Cho are with The Department of Computer Science, Virginia Tech, Falls Church, VA, USA. Email: \{qishengz19, zelin, jicho\}@vt.edu.  Abdullah Zubair Mohammed is with The Bradley  Department of Electrical and Computer Engineering, Virginia Tech, Arlington, VA, USA.  Email: abdullahzubair@vt.edu.  Terrence J. Moore is with US Army Research Laboratory, Adelphi, MD, USA. Email: terrence.j.moore.civ@mail.mil.  The first three authors made almost a same amount of contributions.}}

\maketitle

\pagenumbering{arabic}

\begin{abstract}
Diversity-based security approaches have been studied for several decades since the 1970's.  The concept of {\em diversity-by-design} emerged in the 1980's and, since then, diversity-based system design research has been explored to build more secure and dependable systems. In this work, we are particularly interested in providing an in-depth, comprehensive survey of existing diversity-based approaches, insights, and future work directions for those who want to conduct research on developing secure and dependable cyber-physical systems (CPSs) using diversity as a system design feature.  To be specific, this survey paper provides: (i) The common concept of diversity based on a multidisciplinary study of diversity from nine different fields along with the historical evolution of diversity-by-design for security; (ii) The design principles of diversity-based approaches; (iii) The key benefits and caveats of using diversity-by-design; (iv) The key concerns of CPS environments in introducing diversity-by-design; (v) A variety of existing diversity-based approaches based on five different classifications; (vi) The types of attacks mitigated by existing diversity-based approaches; (vii) The overall trends of evaluation methodologies used in diversity-based approaches, in terms of metrics, datasets, and testbeds; and (viii) The insights, lessons, and gaps identified from this extensive survey.
\end{abstract}


\IEEEpeerreviewmaketitle

\section{Introduction}

\IEEEPARstart{D}{iversity} is an inherent property of the world we live in, which is known as one of key reasons for our survival.  {\em Biodiversity} is well-known as a key factor of the sustainability of an ecosystem in terms of providing proper functionalities, survivability, and even productivity in entities or organizations~\cite{Hong04}.  Inspired by this concept from biodiversity, many diversity-based security mechanisms have been proposed in the literature~\cite{Barrantes03, forrest1997building, Homescu13,  yang2016improving}.  Common examples include different implementations of software providing the same functionalities~\cite{ yang2016improving}, diverse software stacks~\cite{huang2011introducing, huang2010security}, dynamic configurations of a network topology~\cite{zhuang2012simulation}, antenna diversity in hardware for generating a shared secret key~\cite{zeng10}, and architectural diversity to improve security and dependability of Field Programmable Gate Arrays (FPGA) systems~\cite{karam17}.

In this survey, we are particularly interested in investigating how diversity can contribute to enhancing system security and dependability.  Dependability and security are defined by their key attributes, which for dependability includes reliability, availability, safety, integrity, and maintainability while for security encompasses confidentiality, integrity, and availability~\cite{Avizienis04}.  In this work, we conducted an extensive survey on diversity-based approaches designed to build cyber-physical systems (CPSs) more resilient against attacks and faults. To be specific, we focused our survey on the following: the key design principles, the historical evolution of the diversity concept, the key approaches at different system layers, the attacks defended, the evaluation metrics, and the evaluation testbeds used for the proposed diversity-based approaches.  In addition, we extensively illuminate the pros and cons of each approach and address insights and lessons learned from this survey that suggest future research directions.    

To clarify the contributions of our survey paper, we identified the key merits of our survey paper, compared to the existing survey papers discussing diversity-based security approaches~\cite{Balakrishnan05, Baudry15-survey, hosseinzadeh2016, Hosseinzadeh18-survey, Larsen14-survey}.  We conducted a comprehensive discussion that compares our survey paper and the existing survey works on this topic.  Due to space constraints, we provide a brief overview of each survey paper and how it differs from our own with more details provided in Appendix A and the summarization in Table~I of the supplement document.  Unlike the above existing survey works~\cite{Balakrishnan05, Baudry15-survey, hosseinzadeh2016, Hosseinzadeh18-survey, Larsen14-survey}, our survey provided the additional contributions as below. 

The {\bf key contributions} of our survey paper are as follows: \begin{enumerate}
\item We conducted an extensive survey on the multidisciplinary concepts of diversity derived from nine different disciplines to provide an in-depth understanding and merits of diversity to maximize their contributions to achieving system security and dependability. 
\item We provided design principles to develop diversity-based security techniques in terms of what-to-diversify, how-to-diversify, and when-to-diversify as design strategies to enhance system security and dependability.
\item We provided an extensive survey on diversity-based approaches based on a classification of five different layers from the physical environment to human factors in order to comprehensively discuss the core role of each technique and its pros and cons.  In addition, we discussed how the key merit of each technique can contribute to improving system security and dependability.
\item We conducted a comprehensive survey on the set of attacks that have been considered by the existing diversity-based security techniques.  This provides a landscape view of what attacks have been mitigated by diversity, leading to our discussion on what other types of attacks diversity-based approaches may prove fruitful in future research. 
\item We provided an in-depth survey on evaluation methodologies, in terms of metrics, datasets, and testbeds used for experiments conducted to validate the existing diversity-based security approaches.  We also suggested how to improve experimental environments in order to offer more practical help to real world applications in enhancing system security and dependability.  
\item Based on our up-to-date and extensive survey on existing diversity-based approaches and our in-depth discussions of their pros and cons, we offered a list of future research directions that may be highly promising to the design of secure and dependable CPSs.
\end{enumerate}

The rest of this paper is organized as follows:
\begin{itemize}
\item Section~\ref{sec:diversity-concept-design-principles} discusses (i) the common concept of diversity based on the concept of diversity discussed in nine different disciplines; (ii) the evolution of diversity-based security approaches from the 1970's to the 2010's; (iii) the key principles of designing diversity-based approaches to build secure and dependable CPSs, and (iv) the key benefits and caveats of diversity-based CPS designs.
\item Section~\ref{sec:key-properties-DS-CPSs} addresses what types of CPSs we address in this work and the key attributes of security and dependability. 
\item Section~\ref{sec:diversity-cybersecrity-techniques} introduces a variety of existing diversity-based approaches to build secure and dependable CPSs. We discussed the existing diversity-based approaches, in terms of the system layer in which an approach is deployed, covering five layers from physical environments to human-machine interactions, along with the discussions of the pros and cons on each approach.

\item Section~\ref{sec:attacks-countermeasured} surveys what types of attacks are defended by the existing diversity-based approaches.

\item Section~\ref{sec:v-v} provides a survey on how existing diversity-based approaches have been verified and validated in terms of metrics, datasets, and evaluation testbeds used.

\item Section~\ref{sec:discussions} discusses the limitations and lessons learned from this comprehensive survey.

\item Section~\ref{sec:conclusion} concludes the paper by summarizing our key findings and suggesting future work directions.
\end{itemize}

\section{Concepts and Evolution of Diversity-based Security, and Their Design Principles} \label{sec:diversity-concept-design-principles}
The concepts of diversity have been discussed in multiple disciplines and have been applied in various forms in the context of each discipline.  In this section, we discuss the multidisciplinary concepts of diversity and the key benefits and caveats when applying the concept of diversity as a design feature to achieve system dependability and security.
In Appendix B and the summarization in Table II of the supplement document, we surveyed the concepts of diversity from multiple disciplines, including biodiversity, geodiversity, biology, sociology, psychology, political science, organization management, nutrition science, and computation and engineering. Based on this conceptual review of diversity, we derived one common ideology of diversity as follows:

\begin{quote}
{\em Diversity of components in a system (e.g., a group, community, society, body, ecosystem, and computer system or network) can enhance sustainability originated from the principle of polyculture system components that will be highly resistant against sudden, disastrous changes from external effects.  The system sustainability can be achieved by meeting multi-faceted properties of system quality, such as dependability, security, survivability, fault tolerance, resistance, stability, creativity, and resilience.}
\end{quote}




\begin{table*}[t]
\centering
\caption{Evolution of Diversity-based Security and Dependability.}
\label{tab:diversity_evolution}
\vspace{-3mm}
\begin{tabular}{|p{3cm}|p{3cm}|p{3cm}|p{3cm}|p{3cm}|p{3cm}|}
\hline
\cb{\bf 1970's} & \cb{\bf 1980's} & \cb{\bf 1990's} & \cb{\bf 2000's} & \cb{\bf 2010's} \\
\hline
  Emergence of recovery blocks and $N$-version programming (NVP)
  & Enhanced maturity of NVP based on theoretical and empirical analysis; Emergence of `Design diversity’ and `software diversity’
& Combining NVP and `design software’ for software diversity
& Emergence of software diversity for security (e.g., preventing malware)
& More active research on software diversity for security and dependability
\\ 
\hline
\multicolumn{3}{|c|}{\bf Diversity-based software fault tolerance} & \multicolumn{2}{c|}{\bf Diversity-based security and dependability} \\
\hline
\end{tabular}
\vspace{-5mm}
\end{table*}

\subsection{Evolution of Diversity-based Security} \label{subsec:diversity-based-security-applications}

Diversity-based security has been studied for decades.  In the 1970's, \citet{randell1975} proposed recovery blocks in programs to detect potential errors in the execution process and perform spares with diverse implementations as needed.  \citet{avizienis1977implementation} first introduced the concept of $N$-version programming (NVP), providing multiple programmings with the same functionalities.  In the 1980's, \citet{avizienis1985n} and \citet{knight1986experimental} described both experimental results and applications of NVP, which is a fault-tolerance approach that was originally applied to the physical faults and has been reused for software fault-tolerance. \citet{brilliant1989consistent} raised a problem that if the NVP comparison is based on the finite-precision number output from multi-version applications, it is impossible to guarantee that two correct applications have a consistent output leading to potential false positives. 

The terms {\em design diversity}~\cite{avizienis1984fault,avizienis1986dependable} and {\em software diversity} are coined from the hardware diversity~\cite{gmeiner1980software} domain in the 1980's.  In the 1990's, \citet{Cohen93} first applied the concept of diversity in software for defending against cyberattacks.  \citet{forrest1997building} first comprehensively described diversity in computer systems and argued its merit in the application of computer security.  These authors also highlighted the promise of using diversity for security and forecasting some security issues. Also in the 1990's, other studies tried to combine NVP and {\em design diversity}~\cite{lyu1992assuring}. In the 2000's, automate diversity was widely used.  Even still, the purposes of diversity are different even if they all use the same concept of diversity~\cite{Barrantes03}.  For example, in software engineering, diversity is used to create multiple solutions for solving one problem in order to significantly increase the probability of finding a solution. However, in security, diversity is used to avoid replicated attacks and increase attack complexity so the attacker is forced to redesign its strategy even if it attacks the same target.  In the late 2000's, the concept of {\em software diversity} has been applied to defend against malware propagation~\cite{yang2016improving}. An era of the so-called {\em diversity for security} has begun~\cite{chen2005software}.  In the 2010's, software diversity-based approaches were commonly used for enhancing system security and dependability~\cite{Bishop11, crane2015thwarting, Garcia2014}.  We summarized how diversity-based security research has been evolved from the 1970's to the 2010's in Table~\ref{tab:diversity_evolution}.

\subsection{Key Design Principles of Diversity-based Approach} \label{subsec:design-principles}
In this section, 
we discuss key design principles in terms of three aspects: what-to-diversify, how-to-diversify, and when-to-diversify.

\subsubsection{\bf What-to-Diversify}  This principle refers to what platform a given diversity-based approach is applied to achieve a particular design goal.  We discuss the design principle of what-to-diversify for three different systems, including cyber, physical, and cyber-physical systems. What-to-diversify at a different system type is detailed as follows:
\begin{itemize}
\item {\em Diversity-based approaches at cyber systems} have been applied using software stack diversity~\cite{azab2011chameleonsoft, huang2011introducing, jackson2013diversifying}, software version diversity~\cite{alavizadeh2019model, Bishop11, Garcia2014, Gorbenko2019, Totel2005},  code diversity~\cite{bittau2014, Homescu13, hu2006secure, Hosseinzadeh18-survey, Jangda:JITCodeDiversification2015, Koo16}, programming language diversity~\cite{littlewood2001, taguinod2015toward}, or network topology diversity~\cite{zhang2016network}.  The detail of each approach is discussed in Section~\ref{subsec:software}.
\item {\em Diversity-based approaches at hardware systems} have been also used, such as sensors and actuators~\cite{kharchenko16, malynyak18, van2004use}, embedded devices~\cite{gashi2014diversity, wang2019diversity}, and communication modules~\cite{ghourab2017towards, zeng10} to improve the security and dependability of the system. The detail of each approach applied in these categories is given in Section~\ref{subsec:hardware}.
\item {\em Diversity-based approaches at CPSs} have been used  to improve reliability and safety~\cite{kharchenko16}.  The examples using $N$-variance concepts include multi-version technology, multi-version systems, multi-version projects, and multi-version life-cycles.  The application of these concepts on various industrial test-cases have been discussed to improve safety, security, and survivability~\cite{kharchenko16, malynyak18}.
\end{itemize}

\subsubsection{\bf How-to-Diversify} This principle refers to a particular technique to realize diversity in given systems (or networks). We categorize the types of techniques based on the existing approaches as: 
{\em randomization} (e.g., software stack~\cite{azab2011chameleonsoft, jackson2013diversifying}, address space~\cite{bittau2014, Koo16}, instruction set~\cite{hu2006secure}, network shuffling~\cite{temizkan2017software}), 
{\em dynamic reconfiguration} (e.g., code reconfiguration~\cite{Jangda:JITCodeDiversification2015}, reconfiguration of antenna systems~\cite{ghourab2017towards, zeng10}, network topology reconfiguration~\cite{zhuang2012simulation}), {\em diversification} (e.g., software for malware detection~\cite{Bishop11}, the operating system instances~\cite{alavizadeh2019model, Garcia2014, Gorbenko2019}, web-servers~\cite{Totel2005}, code diversity~\cite{Homescu13}, diversified system architecture~\cite{gashi2014diversity, wang2019diversity}), and 
{\em obfuscation} (e.g., code obfuscation~\cite{Hosseinzadeh18-survey}, network diversity~\cite{Neti12}). Each technique is detailed in Section~\ref{sec:diversity-cybersecrity-techniques}.

\subsubsection{\bf When-to-Diversify} Diversity-based approaches can be either dynamically applied (e.g., time-varying dynamic reconfiguration) or statically configured at the system deployment time (e.g., diversification of software stack).  For the dynamic diversification of system configurations, whenever the changes are made, a corresponding cost occurs. Hence, overly frequent changes of system configurations or maintaining too high diversity may introduce some drawbacks. 
Therefore, there should be adaptive strategies that can maintain diversity for system security and dependability while minimizing performance degradation or overhead. 


\subsection{Benefits and Caveats of Diversity-based System Designs}
This section discusses the benefits and caveats of diversity-based approaches to design secure and dependable CPSs.

\subsubsection{\bf The benefits of diversity-based system designs are}
\begin{itemize}
\item {\bf Increasing fault tolerance of a system}:  Diversity-based system design can introduce high fault tolerance, meaning that the system can be functional even in the presence of attacks.   Note that fault tolerance is one of the key attributes of {\em resilience}, which embraces fault tolerance, adaptability, and recoverability~\cite{Cho19-stram}.  The origin of diversity-based approaches was to enhance fault tolerance~\cite{avizienis1984fault}.

\item {\bf Enhancing system availability and reliability}: Software or hardware diversity-based designs allow a system to continuously function even when a system component is being compromised because the system does not consist of homogeneous components exposing the same vulnerabilities.  This introduces high fault tolerance of the system. This also naturally introduces high system availability and reliability since better preparation for attacks increases the capability of providing seamless, uninterrupted services.

\item {\bf Leveraging existing technologies}: A basic approach of diversity-based system design is the use of different implementations of software, hardware, or other system components that can provide the same functionalities or services.  Instead of developing a new technology, which can be challenging as it requires additional time and effort to ensure performance and security requirements, diversity-based design can easily leverage legacy technologies. 
\end{itemize}

\subsubsection{\bf The caveats of diversity-based system designs are} 
\begin{itemize}
\item {\bf Potential high cost and performance degradation}: The key caveat of diversity is the potential of greater cost.  For many types of systems, diversification of system components can be costly to maintain. Hence, even if diversity-based system designs can provide high security and dependability for CPSs, we should seek critical tradeoff-aware, diversity-based designs.

\item {\bf High challenges in deployment of diversity-based designs}: If diversification is not successfully deployed, system security and dependability may suffer since the high cost, delay, or incompatibility may significantly reduce Quality-of-Service (QoS) possibly resulting in system failure due to highly disruptive services. 

\item {\bf Lack of positive effect of diversity in poorly designed, unsecure systems or components:}  Diversity can be effective only when an individual system component is sufficiently secure.  For example, if an individual software package is poorly developed with significant vulnerabilities, using different software packages may not be able to enhance system security or dependability~\cite{Chen18-network-diversity}.

\end{itemize}

\section{Types of CPSs and the Key Attributes of Dependable and Secure CPSs} \label{sec:key-properties-DS-CPSs}
In this section, we discuss three types of CPSs, including Internet-of-Things (IoT), Smart Cities, and Industrial Control Systems (ICSs). In addition, we discuss the dependability and security attributes of CPSs along with how diversity-based designs can enhance the dependability and security of CPSs.

\subsection{Types of Cyber-Physical Systems} \label{subsec:types-CPSs}

\subsubsection{\bf Internet-of-Things} IoT technologies have become more popular and have been recognized as one type of CPS that can provide effective services to users.  IoT encompasses a large number of CPSs including Wireless Sensor Networks (WSNs) integrated with IoT~\cite{kocakulak2017overview}.  {\bf The key challenges of designing an IoT~\cite{Roman2013ComNet}} are: (1) Distributed communications, data filtering/processing, and a large amount of data dissemination in highly different forms (e.g., text, voice, haptics, image, video) for distributed communications in large-scale networks by heterogeneous entities (e.g., devices or humans); (2) resource constraints in battery, computation, communication (e.g., bandwidth), and storage; (3) highly adversarial environments, introducing compromised, deceptive entities and data; and (4) high dynamics of interactions between entities, data, network topology and resources available in which each component dynamically changes in time and space.  


\subsubsection{\bf Smart Cities} From a technological perspective, a smart city is considered as a CPS~\cite{cassandras2016smart}. A smart city encompasses intelligent transportation, smart buildings and infrastructure, smart citizens and governance, and so forth.  In addition to the challenges discussed for IoT, {\bf the following additional design challenges should be considered for smart cities}~\cite{batty2012smart}: (1) Effortless connectivity and coordination of all sectors where the infrastructure of a city is significantly distributed and the smart city should work effectively; (2) Effective and efficient data and their integration required where huge amounts of data need to be generated and systems should perform efficient data acquisition, mining, integration, transformation, and additional analysis; and (3) High security and privacy for data~\cite{elmaghraby2014cyber} where the inherent heterogeneity of smart cities and multiple interfacing systems increases the number of security threats and vulnerabilities~\cite{gharaibeh2017smart}. 

\subsubsection{\bf Industrial Control Systems}
An ICS is an umbrella term for all the control systems associated with industrial processes and instrumentation. As the automation and smart control areas have grown, an ICS deploys a CPS for its purposes. Industrial CPSs are deployed in environments that are not easily accessible by humans and are expected to sustain for a long duration.  {\bf The key challenges of designing an ICS are}: (1) Control timing requirements and complexity where the complex industrial processes have the stringent requirements on timing for efficient, synchronized and uninterrupted operation~\cite{stouffer2011guide}; (2) The nature of distributed environments where a system is distributed across a wide geographical area~\cite{stouffer2011guide}; (3) High availability in which all the domains need to be available at all times~\cite{stouffer2011guide};  (4) Mitigation of a single failure in one domain, which may result in severe consequences on an entire system~\cite{stouffer2011guide}; and (5) High vulnerabilities to cyberattacks~\cite{drias2015analysis} caused by the large attack surface derived from the distributed nature of the ICS, the system requiring user or device authentication (e.g., a two-way authentication), and communication channels vulnerable to eavesdropping or DoS attacks.

\subsection{Key Attributes of Dependability and Security for CPSs} \label{subsec:DS-attributes}
In this section, we discuss the key attributes of system dependability and security and how diversity-based designs can contribute to build secure and dependable a CPS.

\subsubsection{\bf Dependability Attributes for CPSs} \citet{Avizienis04} defined the primary attributes of dependability as availability, reliability, safety, integrity and maintainability. They discussed robustness as a secondary attribute of dependability to be examined against external faults. However, \citet{Cho19-stram} argued that resilience and agility should be also considered as dependability attributes since resilience and agility can capture the dynamic aspects of a system, which have not been fully addressed in the other existing metrics.  According to \citet{iso92}, dependability represents a collective term used to describe system availability and its influencing factors, including reliability, maintainability, and maintenance.

\subsubsection{\bf Security Attributes for CPSs}
According to \citet{Avizienis04}, the primary security attributes consist of confidentiality, integrity, and availability.  The secondary security attributes include accountability, authenticity, and non-repudiation.  \citet{kharchenko16} included availability, confidentiality and integrity as security attributes.  
\citet{humayed2017} studied vulnerabilities and attacks in smart grids, medical CPS, and smart cars, where they interpreted security as availability, one of the security goals.  \citet{Trivedi09} defined dependability and security as one property where their attributes are defined based on availability, confidentiality, integrity, performance, reliability, survivability, safety, and maintenability.  Compared to~\cite{Avizienis04}, performance and survivability are additionally considered in~\cite{Trivedi09}.

\subsubsection{\bf Discussions -- How Diversity-based Design Can Enhance Dependability and Security?} Due to multiple versions of systems with the same functionality, diversity-based designs can enhance dependability attributes in all aspects, including availability, reliability, safety, integrity, and maintainability. However, diversity can have multiple, different effects on security attributes, such as confidentiality, integrity, and availability. For example, high diversity design may enhance system availability and integrity while it may not necessarily increase confidentiality.  

\section{Diversity-based Techniques} \label{sec:diversity-cybersecrity-techniques}
In this section, we discuss the existing diversity-based techniques for dependable and secure CPSs.  In order to embrace multi-faceted aspects of a CPS, we categorize diversity-based techniques at the following layers: Physical environments, network, hardware, software, and human users. 

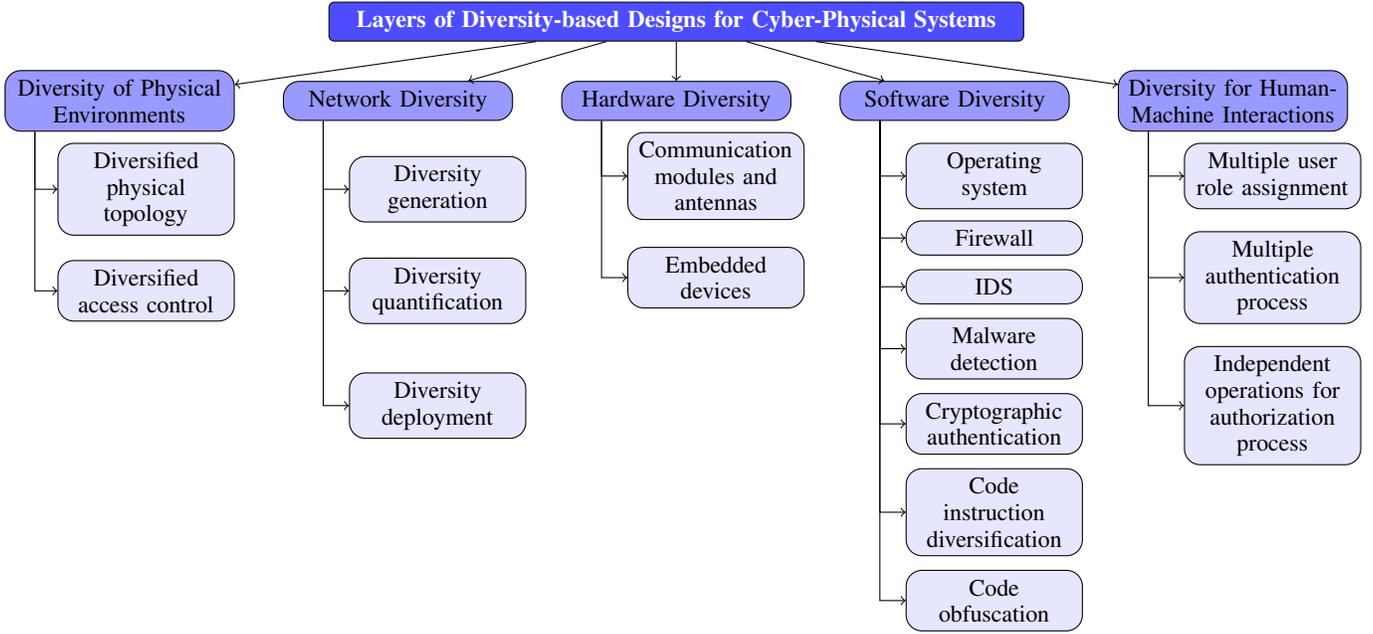
\begin{figure*}
\centering
\begin{tikzpicture}[
    basic/.style  = {draw, text width=9cm, rectangle, font=\small},
    root/.style   = {basic, rounded corners=2pt, thin, align=center, fill=blue!70},
    level 1/.style={basic, rounded corners=6pt, thin,align=center, fill=blue!40, text width=8em, sibling distance=37mm},
    level 2/.style = {basic, rounded corners=6pt, thin, align=center, fill=red!60, text width=5em, sibling distance=28mm},
    edge from parent/.style={->,draw},
    level 3/.style = {basic, rounded corners=6pt, thin, align=center, fill=blue!10, text width=6em},
    level distance=3em
  ]
  
\node[root] {\textcolor{white}{\bf Layers of Diversity-based Designs for Cyber-Physical Systems}}
  child {node[level 1] (c1) {Diversity of Physical Environments}}
  child {node[level 1] (c2) {Network Diversity}}
  child {node[level 1] (c3) {Hardware Diversity}}
  child {node[level 1] (c4) {Software Diversity}}
  child {node[level 1] (c5) {Diversity for Human-Machine Interactions}};

\begin{scope}[every node/.style={level 3}]
\node [below of = c1, xshift=10pt, yshift=-5pt] (c11) {Diversified physical topology};
\node [below of = c11, yshift=-10pt] (c12) {Diversified access control};
\node [below of = c2, xshift=15pt, yshift=-5pt] (c21) {Diversity generation};
\node [below of = c21, yshift=-10pt] (c22) {Diversity quantification};
\node [below of = c22, yshift=-15pt] (c23) {Diversity deployment};
\node [below of = c3, xshift=15pt] (c31) {Communication modules and antennas};
\node [below of = c31, yshift=-10pt] (c32) {Embedded devices};
\node [below of = c4, xshift=15pt] (c41) {Operating system};
\node [below of = c41, yshift=5pt] (c42) {Firewall};
\node [below of = c42, yshift=10pt] (c43) {IDS};
\node [below of = c43, yshift=5pt] (c44) {Malware detection};
\node [below of = c44, yshift=0pt] (c45) {Cryptographic authentication};
\node [below of = c45, yshift=-5pt] (c46) {Code instruction diversification};
\node [below of = c46, yshift=-5pt] (c47) {Code obfuscation};
\node [below of = c5, xshift=15pt] (c51) {Multiple user role assignment};
\node [below of = c51, yshift=-10pt] (c52) {Multiple authentication process};
\node [below of = c52, yshift=-20pt] (c53) {Independent operations for authorization process};
\end{scope}


\foreach \value in {1,...,2}
  \draw[->] (c1.200) |- (c1\value.west);

\foreach \value in {1,...,3}
  \draw[->] (c2.195) |- (c2\value.west);

\foreach \value in {1,...,2}
  \draw[->] (c3.195) |- (c3\value.west);
  
\foreach \value in {1,...,7}
  \draw[->] (c4.195) |- (c4\value.west);

\foreach \value in {1,...,3}
  \draw[->] (c5.200) |- (c5\value.west);

\end{tikzpicture}
    
\caption{Layers of Diversity-based Designs for Cyber-Physical Systems.}
\label{fig:multi_layer_CPS}
\vspace{-5mm}
\end{figure*}


\subsection{Diversity of Physical Environments} \label{subsec:phy-env}
A CPS incorporates a physical environment to further improve its practicality and effectiveness. These physical environmental factors include guards, cameras, badge readers, physical access policies, biometrics, and so forth~\cite{weingart00}. However, this may introduce different types of vulnerabilities in CPSs, implying a widened attack surface.  \citet{skandhakumar12} proposed a building information model to capture the diversified topology information of the building and construct a 3D model of the building. This information can be used to administrate and enforce the access control systems. \citet{akhuseyinoglu17} built a general access control model to manage potential risks in access requests.  For any incoming accepted access request, their model could check multiple diversified action sequences to select the option that can minimize risks to the system. \citet{cao20} developed a risk control mechanism to dynamically assign clearance based on a users' history of activities and the related risks of the requests. They considered diversified physical topology constraints to further restrict the access to important assets in CPSs.  Many physical environment security issues are related to the human factors because humans can introduce serious security vulnerabilities during their interaction with physical systems~\cite{patrick03}. We discuss more on this in Section~\ref{subsec:human-factors}.

\noindent {\bf Pros and Cons}:
Diversification in physical environments is generally more complicated than that in cyberspaces due to the high uncertainty
and inconstancy associated with human mistakes, including human attackers that have high intelligence and ability to launch sophisticated attacks. Access control in physical environments is primarily used to manage human-related risk. However, current approaches still introduce high vulnerabilities to highly intelligent human attackers.

\subsection{Network Diversity} \label{subsec:networks}

Typically, a CPS includes many networks, such as sensor networks and actuator networks, where network topology is defined as the set of connections between network components. The network diversity research has been explored primarily in terms of three aspects: diversity generation, diversity quantification, and diversity deployment. We discuss each aspect in detail below.

\subsubsection{\bf{Diversity Generation}}
Network diversity refers to the diversification in network settings. Key factors in network settings include network topology and system components installed in each node of the network.  The system component includes hardware or software components. Thus, network diversity can be derived from heterogeneous network topologies or generated from different variants of hardware and software components, which will be detailed in Sections~\ref{subsec:hardware} and \ref{subsec:software}.

\subsubsection{\bf{Diversity Quantification}}\label{metrics_section}
Diversity metrics have been proposed to measure the degree of network diversity by deploying multiple variants of software. The diversity metrics used in the literature are as follows:
\begin{itemize}
\item {\bf Entropy}: This metric is used extensively to measure the randomness or unpredictability in a system. Entropy is used as an indicator of the richness of species in the field of biodiversity~\cite{Neti12}. The same concept is used to measure diversity in a network where higher entropy indicates higher network diversity, which is assumed to be more secure. 

\vspace{1mm}
\noindent {\bf Pros and Cons}:
Entropy is a common metric to measure the extent of the polyculture of software or hardware to achieve network diversity.  However, entropy-based metrics may not be an effective measure of diversity if the variants share vulnerability to the same attack. In this case, a network with high entropy can even introduce higher vulnerabilities to identical attacks.  In addition, entropy does not measure topological network diversity.

\item {\bf Resilience metrics}: \citet{zhang2016network} devised three diversity-based metrics (i.e., $d_1$, $d_2$, $d_3$) to evaluate the resilience of a network in the presence of software diversity as follows: (1) $d_1$ is a biodiversity-based model, using the number of distinct resources, distribution of resources, and a variety measure of resources for evaluation. It is defined as the ratio of the network effective richness to the total number of variants; (2) $d_2$ evaluates the least attack effort to compromise hosts. It is defined as the ratio of the number of resources in attack path to the number of steps in the attack path; and (3) $d_3$ is similar to $d_2$, but it focuses on the average attack effort. It is a probabilistic model defined by the ratio of the probability of given asset being attacked over the probability of a given asset being attacked with the condition of all variants being unique.

\vspace{1mm}
\noindent {\bf Pros and Cons}: Similar to entropy-based metrics, the $d_1$ metric works well assuming that variants are alike. The limitation of this metric is that it doesn't consider the causal relation between variants. The $d_2$ metric considers the causal relation between variants, but does not focus on enhancing security. It is also computationally costly to evaluate, but the cost can be reduced by estimating its value using heuristics.  The $d_3$ metric provides a global view in terms of average vulnerabilities. Hence, it may not be able to distinguish between two networks with similar average vulnerabilities.
\end{itemize}

\subsubsection{\bf Diversity Deployment} \label{subsubsec:diversity-deployment} Diversity can be further enhanced depending on how the existing diversity can be differently deployed. We call this {\em diversity deployment} and classify these types into two classes: metric-based and metric-free. We discuss each class as below. 

\begin{itemize}
\item {\bf Metric-based Diversity Deployment}: This class provides metrics to measure diversity in a proposed algorithms. For example, \citet{temizkan2017software} proposed a software allocation model using Shannon entropy as a diversity metric to minimize the total information gain in the software assignment problem. The main drawback of this metric is high complexity, as the problem is NP-hard. The authors proposed a heuristic algorithm to reduce the complexity. 

\citet{borbor2019optimizing} leveraged the software diversity metrics defined in \cite{zhang2016network} (i.e., $d_1$, $d_2$, $d_3$ metrics) to present their model-based technique.  This work solved a software assignment problem aiming to optimize a max-min of $d_1$ and $d_2$ and a min-max of $d_3$ among all nodes in a network. This work leveraged a meta heuristic (i.e., a genetic algorithm) to solve the problem.

\vspace{1mm}
\noindent {\bf Pros and Cons}: The use of metrics to measure diversity can provide a simple solution via maximizing the metric assuming that high diversity enhances network security.  If the validity of a diversity metric does not hold, however, the relationship between network diversity and network security may not hold as well.  In addition, high diversity can introduce high overhead as well as potential performance degradation (e.g., incompatibility between nodes).  
Moreover, solving an optimization problem using a diversity metric may have high complexity. Heuristics introduced to solve high complexity, such as meta heuristics, could be also computationally prohibitive in reaching an optimal solution.


\item {\bf Metric-free Diversity Deployment}: This deployment class does not use any metrics to measure diversity; rather it simply uses randomization or dynamic reconfiguration.  To discuss metric-free diversity deployment in detail, we further classify this class into the following two sub-classes: 
\begin{itemize}
\item {\bf Graph Coloring}: This approach, borrowed from graph theory~\cite{jensen2011graph}, seeks to color a graph such that every pairwise connected nodes have different colors.  This idea is reformulated into its variant in the domain of software diversity. Different software variants, representing different colors, are expected to be installed into each node in a computer communication network. Therefore, coloring techniques aim to assign different software versions to every pair of connected nodes. Leveraging this idea, a software assignment problem is solved by~\citet{o2004achieving} by developing different coloring algorithms.  In addition, \citet{huang2017software} studied the order of coloring based on priority determined using different centrality metrics. Taking this approach further, \citet{touhiduzzaman2018diversity} introduced a game theoretic approach to solve a graph coloring problem by using different software versions to minimize vulnerabilities to epidemic attacks. 


\vspace{1mm}
\noindent {\bf Pros and Cons}: Since the coloring problem has been studied for decades in mathematics and other engineering domains, its theoretical validity and maturity for algorithmic effectiveness and efficiency has been already proven and can be reliably leveraged.  However, as the coloring problem has predominantly been studied in static networks, its applicability in dynamic network that can ensure efficiency as well as effectiveness is not fully proven. Moreover, simple repetition of the static network-based coloring algorithm may introduce high reconfiguration overhead.

\item {\bf Network Topology Shuffling}: This technique aims to identify an optimal assignment of software variants to maximize the degree of software variants along attack paths.  The main idea is to increase attack cost or complexity for an attacker by increasing hurdles in reaching a target node.  
\citet{hong2017optimal} solved a network shuffling problem for software assignment as an online moving target defense.  Their proposed algorithm is designed to redirect a certain number of edges for reconfiguring a network topology to be robust against worm attacks.

\vspace{1mm}
\noindent {\bf Pros and Cons}:
Shuffling techniques can cope with dynamic network structures because the cost of redirecting edges is relatively low.  Furthermore, network shuffling does not require assigning software variants. However, the complexity of network shuffling algorithms is proportional to the number of edges in a network, which may not scale for large networks. 
In addition, if a network is required to stay in the same network topology, network shuffling may not be applicable.
\end{itemize}
\end{itemize}

To introduce diversity into a system, existing dynamic reconfigurability techniques are often leveraged.  Dynamic reconfigurability refers to the ability to dynamically reconfigure system settings, such as network topology and software resources. The reconfiguration process may introduce various types of diversification. For instance, dynamically reconfiguring network topology would bring path diversity while software resources reallocation would result in software diversity.  To clarify the scope of our survey paper, we treat dynamic reconfiguration as a subset of diversity-based solutions because diversity-based approaches can be also applied in static network environments. 

\subsection{Hardware Diversity} \label{subsec:hardware}

The hardware of a CPS constitutes sensors and actuators, communication modules and antennas, and embedded devices~\cite{regazzoni17}.  The sensors and actuators form a bridge between the \textit{cyber} and the \textit{physical} parts. In sensors, their physical information is translated to electrical voltages or currents (the opposite in case of the actuators), usually in the order of milli-volts or milli-amperes. These components are vulnerable to false data injection, generally in the form of intentional electro-magnetic interference (IEMI)~\cite{kune2013ghost, selvaraj2018}. To the best of our knowledge, diversity-based security techniques have not been studied yet for sensors and actuators. Therefore, in this section, we only discuss diversity-based techniques proposed for communication modules and antennas and embedded devices.

\subsubsection{\bf Communication Modules and Antennas} In a CPS, the communication module is responsible for the transmission and reception of information (i.e., control and data) between nodes. The medium of communication can be wired, for example, in IEEE 802.3 (Ethernet) standard or wireless, as in standardsIEEE 802.11 (WiFi), IEEE 802.15.1 (Bluetooth), or IEEE 802.15.4 (Zigbee). In a wireless medium, diversity is usually achieved by employing multiple antennas for communication. From the physical layer security perspective, diversity is used to achieve high {\it secrecy capacity} or low {\it intercept probability}. These metrics quantify the ability of a wireless channel to protect its data from a malicious eavesdropper.

\citet{zou2015improving} discussed the effects of diversity on the physical layer security of a communication system, with the following three types of diversity: (1) {\em Multiple input multiple output} (MIMO) diversity where multiple antennas are used for transmitting and receiving nodes; (2) {\em Cooperative diversity} in which multiple relays (repeaters) are used between the transmitting and the receiving node; and (3) {\em Multiuser diversity}, where the transmitter node is communicating with multiple receiver nodes. Via experiments, the authors showed that diversity of the communication channel introduces higher secrecy capacity and lower intercept probability.

\citet{ghourab2017towards} used multiple antennas to reconfigure the frequency of the transmitter node periodically to improve the channel secrecy capacity and enhance the performance against an eavesdropping attack on the routing path in the network. The authors further enhanced the security by obfuscating the transmitted data by intentional injection of false data for diversifying both in space and time.  
\citet{watteyne2009reliability} proposed that the transmitting node sends the succeeding packets on different frequencies, introducing frequency hopping in routing. The system is protected against communication failures in the path, considerably improving the reliability.  \citet{zeng10} introduced a key generation protocol with the contribution of generating high speed key generation and security against passive eavesdropping attacks.  \citet{sarkar2012enhancing} employed the channel diversity to improve the secrecy capacity against an eavesdropper.

\vspace{1mm}
\noindent {\bf Pros and Cons}: Diversity-based security has been mainly developed by diversifying channel frequency, which has been proven highly effective for enhancing system security.  However, most of the approaches require multiple antennas at the transmitter and receiver that increases hardware complexity and requires non-trivial signal processing. This leads to high power consumption, thus requiring lightweight solutions to realize diversity of channel frequency.

\subsubsection{\bf Embedded Devices}
An embedded device, including microcontrollers, microprocessors, FPGAs or Application Specific Integrated Circuits (ASICs), is the core of a CPS. The embedded device acts as a central controller of the system as well as provides an interface between the cyber and physical aspects of the system. We discuss diversity-based designs in embedded devices as follows:
\begin{itemize}
\item {\bf Architectural Diversity of FPGAs}: \citet{lach1999algorithms} exploited the intrinsic redundancy and re-configurability in the architecture of FPGAs to improve its fault tolerance and reliability.  \citet{karam17} used the architectural diversity in FPGAs to generate a different final executable file (i.e., bitstream) in each of the nodes in the network. This increases difficulty in reverse engineering techniques performed by attacker, leading to enhanced security against tampering and piracy attacks with reasonable overhead. 

\item {\bf Variants of Physical Layer Identification}: Even if devices are of the same model from the same manufacturer, there exists minor variations in the intrinsic characteristics at their physical layer, such as transients in the radio signals, clock skew, and other features.  This diversity can be utilized for physical layer identification (PLI) and device fingerprinting as a security technique defending against impersonation and identity-theft attacks~\cite{gerdes2011physical}.  \citet{danev2009transient} identified the IEEE 802.15.4 devices using the variations in the turn-on transients of their radio transceivers.  \citet{gerdes2006device} used a matched-filter based approach to identify Ethernet devices from the variations in their analog signal.  \citet{foruhandeh2019simple} proposed a technique to identify Electronic Control Units (ECUs) in a Controller Area Network (CAN) bus architecture by exploiting the variations in their physical layer features.  \citet{cobb2010physical} introduced radio frequency distinct native attribute (RF-DNA) fingerprinting to identify embedded processors from their RF emissions to detect intrusions and prevent against impersonation attacks.


\vspace{1mm}
\noindent {\bf Pros and Cons}: This inherent variation between devices increases their resiliency against side-channel attacks. However, side-channel attacks are less effective if an attacker is trained on one device and tested on another device from the same manufacture and with an identical chip~\cite{wang2019diversity}.  The attacker can improve its efficacy by training on a diverse set of devices and on varied implementations of the cryptographic algorithm.  In addition, modulation-based identification as a PLI method may be vulnerable to signal and feature replay attacks while transient-based identification is more robust~\cite{danev2009transient}.  Hence, it is important to appropriately use a relevant diversity design to deal with the given attack scenario.
\end{itemize}

\subsection{Software Diversity} \label{subsec:software}
Software diversity-based approaches have been substantially used to enhance system security as the key diversity-based design.  Due to the large volume of studies explored in the literature, we mainly looked at the following different types of diversities applied in: (i) Operating systems; (ii) firewalls; (iii) intrusion detection systems (IDSs); (iv) malware detection; (v) cryptographic authentication; (vi) instruction diversification; and (vii) code obfuscation.   

\subsubsection{\bf Operating Systems (OS)}  \citet{forrest1997building} provided the following guidelines for designing OS-based diversity methods to preserve their convenience, usability and efficiency: (i) preserve a high-level functionality; (ii) introduce diversity that can disrupt known intrusion most; and (iii) minimize deployment cost and run-time cost while maintaining sufficient diversity.  \citet{Garcia2014} also proved the usefulness of OS diversity by analyzing the vulnerabilities in 11 different OSs collected over 15 years. The analysis showed that many vulnerabilities exist in more than one OS; but if several OSs are combined, the number of common vulnerability decreases.  They also proposed a method to identify optimal composition of diverse OSs by analyzing the data from the National Institute of Standards and Technology (NIST) National Vulnerability Database (NVD) to improve intrusion tolerance. Based on the common vulnerability (CV) that refers to the same vulnerability found in more than one system, they proved that the OS diversity protects a system from attacks aiming to penetrate into a given system.  

\citet{Gorbenko2019} designed an optimal intrusion-tolerant architecture composed of several different OSs in which a request will pass through these OSs synchronously. If the responses from these OSs are not the same, then an intrusion may happen. 
The authors showed that a 3-variant system is an optimum configuration providing the least vulnerabilities in availability and integrity.   \citet{Garcia2014} found that reducing the number of days of gray-risk and the number of forever-day vulnerabilities are main challenges for OS security. In addition, if a large section of code is reused from a previous version, buffer overflow vulnerabilities may remain in even new technologies proposed to deal with such vulnerabilities.

\citet{nagy2006n} applied an $N$-version technique on an OWASP (Open Web Application Security Project) to enhance the robustness against common vulnerabilities.  In addition, \citet{nagy2006n} further discussed the feasibility of detecting zero-day attacks.  
\citet{Calton96} developed a specialization toolkit for helping a programmer to improve the resistance of systematic specialization of OS kernels and against virus and worm attacks. The toolkit protects an OS by dynamically generating various versions of software components at compile-time specialization and run-time specialization. 

\vspace{1mm}
\noindent {\bf Pros and Cons}: OS diversity techniques are high-level strategies in software architecture, which allow them to defend an attacker while having a certain number of code errors. However, there always exists a trade-off between the cost and diversity, and this conflict becomes more apparent in this category since most of the techniques require multiple OSs to run in parallel. 

\subsubsection{\bf Firewalls}
\citet{liu2008diverse} proposed a process of designing diverse firewalls for enterprise security based on three processes: design, comparison, and resolution.  The {\em design phase} lets multiple groups design a firewall policy independently based on the same requirement.  The {\em comparison phase} detects function discrepancies between the multiple policies. The {\em resolution phase} generates a unified design for all groups.  The authors also proposed three algorithms to identify all functional discrepancies and estimate the impact of a policy change at the comparison phase. Based on~\cite{Liu04}, the authors also conducted a firewall policy impact analysis~\cite{liu2007change}.

\vspace{1mm}
\noindent {\bf Pros and Cons}: Diversity-based security mechanisms in firewalls are known to be very effective to deal with zero-day attacks~\cite{liu2008diverse}. However, if diverse software is configured or designed by the same group of people, they may share a common problem, which eliminates the advantage of diversity~\cite{weber1999firewall}. In addition, the research on diversity-based firewall to enhance system security is still in its infancy, showing a lack of studies in this research area.  This could be because of the overhead and potential errors introduced due to continuous firewall policy changes. Cost-effective firewall policy changes should be considered for security enhancement.


\subsubsection{\bf Intrusion Detection Systems (IDSs)} \citet{Reynolds2002} proposed an implementation for protecting users of a web service from cyberattacks. This approach exploits commercial off-the-shelf (COTS)-based diversity for IDSs and provides a partial resolution to detect and isolate network attacks.  \citet{Totel2005} utilized the diversity  of COTS to build an IDS and evaluated their proposed IDS at the web-server level using three different servers, namely, Buggy HTTP (Linux), IIS (Windows), and Apache (MacOS). The IDS was tested against seven different types of attacks from the CVE~\cite{cve}.  The IDS detected all the attacks against one of the servers. \citet{reynolds2003line2, reynolds2003line} proposed a diversity-based system to positively identify attackers with `sandboxes'. This system is designed for protecting web servers from on-line intrusion by comparing outputs of diverse software.

\citet{Qu2018} exploited diversity in the implementation of web applications to develop a technique to defend against code injection attacks. They evaluated 16 web applications written in four diverse languages, PHP, ASP, ASP.NET and JSP against SQL injection vulnerabilities from the CVE~\cite{cve}.  Their results showed that the proposed approach has 0\% False Positive Rate (FPR), 25.93\% False Negative Rate (FNR) and 98.03\% detection accuracy. All these results clearly exceed the single-stage counterpart.  \citet{cox2006n} provided an architectural framework to detect and disrupt large classes of attacks.  The framework contains a polygrapher to receive input and copy to a different server exhibiting anomaly behavior. 

\citet{gu2008principled} proposed a `decision-theoretic alert fusion technique' to deal with the alarms from multiple IDSs. This technique is based on the likelihood ratio test (LRT) to combine different alert reports. In addition, since there is little work on analyzing the effectiveness of an IDS ensemble, this technique evaluates the effectiveness of the IDS ensemble by testing the LRT rule on two different datasets in advance.  \citet{majorczyk2007experiments} provided a `masking mechanism' for an IDS to resolve the high FPR when applying COTS-based diversity. Instead of directly assigning a request to diverse components and comparing outputs, the masking function can modify the request before and after the request being processed by diverse components. 

\vspace{1mm}
\noindent {\bf Pros and Cons}:
IDS research is another field mainly studied within Enterprise Security or Web Server Security in which COTS diversity is used by many projects. The combination of various detectors is a well-known strategy that can achieve better performance. However, one main issue associated with the COTS-based IDS is that the FPR can increase due to the types of detection algorithms used in each detector and the ways of estimating diversity across different detectors. How to decrease the FPR while increasing the use of diversity-based IDSs is a promising direction~\cite{majorczyk2007experiments}. 

\subsubsection{\bf Malware Detection}
\label{subsubsec:malware}
\citet{Oberheide2008} conducted 7,220 unique malware tests based on the datasets from NetScout systems~\cite{netscout} and compared the effectiveness of a single malware detector and an ensemble of multiple malware detectors.
\citet{Hole2013} provided a diversity software for an enterprise networked computer system to slow down or prevent the spreading of infectious malware and prevent zero-day exploits. 

Leveraging the benefits of using diverse Antivirus (AV) software, \citet{Silva2010} developed an AV system for the e-mail framework using different AVs running in parallel.  They evaluated their system on e-mails containing malware.  \citet{gashi2009experimental} investigated the benefits of diverse COTS antivirus by analyzing 1,599 malware samples for 178 days~\cite{leita2008sgnet_2}, and with 32 different AV products. 
\citet{Bishop11} reviewed this same dataset to study the detection gains when utilizing more than two AV products as well as to demonstrate the reduction in `at risk time' of the system. 
\citet{smutz2016tree} provided a diversity-based evasion detection method to improve the evasion resistance of a malware detector. 

\vspace{1mm}
\noindent {\bf Pros and Cons}: Diversity-based malware detectors are known to be very effective in defending against zero-day attacks, compared to other security technologies, such as anti-malware or patching. However, their downside is the consumption of more computing resource since different detectors are required to work in parallel, such as Silva's diversity-based antivirus software~\cite{Silva2010}. 

\subsubsection{\bf Cryptographic Authentication} 

\citet{Carvalho-msthesis14} proposed a redundancy and diversity-based method for cloud authentication resistant against unknown, zero-day vulnerabilities.  The key idea is to use redundant authentications to ensure reliability while diverse authentications are used for fault tolerance under attacks compromising part of the system.  

\noindent {\bf Pros and Cons}: In the literature, diversity-based cryptographic authentication has been rarely studied as we only cited one work above~\cite{Carvalho-msthesis14}.  The main reason of lack of studies in this area would be because using multiple authentication protocols may introduce more complexity in system performance as well as incompatibility with other systems that use different authentication mechanisms. 

\subsubsection{\bf Code Instruction Diversification} 
This technique is to diversify code instructions to prevent side-channel attacks, code modification attacks, or code replay attacks.

\citet{Homescu13} created various programs by randomly inserting NOP (No Operation Performed) instructions prior to compiling them.  An NOP instruction does nothing but is used for randomizing the code layout. 
\citet{Koo16} used an instruction displacement to randomize the starting addresses of gadgets in the binary on installation phase. 
\citet{Williams09} proposed a technique, called {\em Calling Sequence Diversity} (CSD) where the call sequence is defined as the sequence of instructions for call and return behavior~\cite{johnson1981computing}.  
\citet{Kc03} developed new randomized instruction sets for each process when the program is loaded to the main memory.  \citet{Barrantes03} proposed a Randomized Instruction Set Emulation (RISE) based on the Valgrind binary translator. 

\citet{hu2006secure} and \citet{Williams09} improved the performance of the Instruction Set Randomization (ISR), a technique for randomly altering instructions~\cite{sovarel2005s}.  \citet{hu2006secure} combined the ISR with the Advanced Encryption Standard (AES) to operate a dynamic translation to software and to improve the efficiency of the ISR. \citet{Williams09} further improved the efficiency of the ISR by leveraging the extended tool chain and combining static and dynamic binary rewriting.  \citet{Franz10} proposed a mechanism for the {\em App Store} to  automatically compile out identical binary code for different devices. 
\citet{Cohen93} proposed a method that can make a static program self-evolve over time for increasing attack complexity based on the {\em Instruction Equivalence} and {\em Instruction Reordering} technologies.  \citet{Chew02} proposed lightweight methods for mitigating buffer overflow by randomizing system call mapping, global library entry point, and stack placement.  \citet{Xu03} proposed a Transparent Runtime Randomization (TRR) method to defend against a wide range of attacks. The TRR, implemented by a program loader, dynamically relocates stack, heap, shared libraries in the memory space of a program.  \citet{ichikawa2008diversification} developed a diversified instruction set architecture (ISA) to increase the redundancy of software. The key idea of the ISA is to change the encoding of opcode while keeping the original functionality of an instruction set.

\vspace{1mm}
\noindent {\bf Pros and Cons}:  Along with OS diversification, code instruction diversification research has been substantially explored.  The key reason would be its less adverse impact on system performance while maintaining the original functionality of the code.  However, it is inevitable that code instruction diversification introduces the complexity of the coding process and incurs high CPU overhead. 

\subsubsection{\bf Code Obfuscation}  
This technique aims to transform code and make it unintelligible but still functional.

\citet{collberg1998manufacturing} designed a code obfuscator for Java that inserts opaque predicates into a Java program and generates an equivalent one but harder to reverse engineer. 
\citet{kuang2016exploiting} and \citet{kuang2018enhance} enhanced the existing VM-based code obfuscation, such as the Code Virtualizer~\cite{CodeVirtua} and VMProtect~\cite{VMProtect} by adding a dynamic instruction scheduler to randomly direct a program. This new approach is called {\em the dynamic scheduling for VM-based code protection} (DSVMP), which was designed to increase the robustness of code obfuscation against highly intelligent attackers capable of using obfuscation techniques. 

\citet{xue2018exploiting} proposed an obfuscation scheme, called {\em Code Virtualization Protection with Diversity} (DCVP), to increase complexity even for experienced attackers to uncover the virtual instructions to native code when applying code virtualization for code obfuscation.  The underlying idea of DCVP is to obfuscate the mapping between the opcodes and semantics for increasing the diversity of the program behavior.  \citet{pawlowski2016probfuscation} proposed an obfuscation technique based on probabilistic control flow. The key idea is generating different, multiple execution traces while keeping semantics, given the same input values. Via experiments, the authors proved that their developed obfuscation prototype can effectively ensure divergent traces for the same input while enhancing resilience under dynamic attack analysis.  
\citet{crane2015thwarting} proposed a dynamic control-flow diversity technique to defend against online and off-line side-channel attacks. 
The authors proved that this technique protects the program under cache-based side-channel attacks aiming to obtain a cryptographic key by analyzing the program execution.  \citet{hataba2015diversified} proposed a technique to dynamically disrupt the control flow of a program so that the conditional branches will be converted randomly. This method is designed for mitigating the side-channel attack in a cloud platform.

\vspace{1mm}
\noindent {\bf Pros and Cons}: Even though the code obfuscation is designed to transform code and make it unintelligible but still functional, the obfuscation cannot guarantee the irreversibility of the code. However, it can still increase the cost for the attacker to understand the functionality of the code, which can increase the opportunity time to protect the program~\cite{Hosseinzadeh18-survey}. A well-known drawback is that many technologies involved in running VMs in parallel require tremendous computational resources.



\subsection{Diversity for Human-Machine Interactions} \label{subsec:human-factors}
Human-machine interface deficiencies are commonly caused by interface design faults that can introduce data delay display or data misinterpretation.  In addition, a single authorization mechanism, such as password only or biometric only, can expose security vulnerabilities. To enhance the security of an authorization system, diversity-based designs can be introduced~\cite{clark1987comparison, reiter1996distributing}. For example, \citet{clark1987comparison} proposed a user level diversity strategy called the `separation of duty' for military security systems.  This strategy assigns the complementary roles to different users and makes the sensitive operation executable with different roles.  \citet{reiter1996distributing} developed a protocol to force a sensitive operation for authorization to be run on different machines or programs controlled by independent operators.

\vspace{1mm}
\noindent {\bf Pros and Cons}: Increasing diversity of human-machine interfaces can increase system security. However, due to humans' limited cognition, high diversity of the human-machine interfaces may introduce more mistakes or errors by humans.  \citet{huang2014links} investigated how human error diversity is related to software diversity under various conditions. Depending on the human operators' skill levels, the human error diversity is shown differently under a different level of software diversity. However, regardless of the skill levels of the human operators, the design of software diversity should be considered human-friendly to minimize human-prone mistakes or errors.  \citet{deswarte1998diversity} proposed a principle to eliminate errors by requiring several independent operators to perform sensitive operations. However, we still need some metrics to estimate the extent of diversity of human operators and sensitivity values of the operation.

Fig.~\ref{fig:multi_layer_CPS} shows an overview multi-layered structure of the classification.  Table~III in the supplement document (Appendix C) summarized the overview of the existing diversity-based approaches surveyed in this section. 

\section{Attack Types Considered by Diversity-based Security Approaches} \label{sec:attacks-countermeasured}

In this section, we mainly discuss what types of cyberattacks are defended by diversity-based security solutions. The limitations and gaps identified from the existing attack model are discussed in Section~\ref{sec:discussions} along with other limitations and insights learned from this survey paper.  
\begin{figure}[h]
\vspace{-3mm}
\centering
\begin{tikzpicture}
[pie chart,
slice type={Physical attack}{blue!90},
slice type={Zero-day attack}{blue!70},
slice type={Worm attack}{blue!50},
slice type={Code injection attack}{blue!30},
slice type={Code reuse attack}{blue!10},
slice type={Return-to-Libc attack}{orange!30},
slice type={Correlated attack}{orange!50},
slice type={Coordinated attack}{orange!70},
slice type={Buffer overflow attack}{brown!30},
slice type={Side-channel attack}{brown!50},
slice type={Deobfuscation attack}{Black},
slice type={Impersonation attack}{Black!90},
slice type={Tampering attack}{Black!70},
slice type={Eavesdropping attack}{Black!50},
slice type={Denial of Service attack}{Black!30},
slice type={Reverse engineering}{Black!10},
pie values/.style={font={\small}},
scale=2]
\pye[xshift=0.6cm,values of coltello/.style={pos=1.1}]{1994}{
3/Physical attack,
12/Zero-day attack,
14/Worm attack,
7/Code injection attack,
3/Code reuse attack,
1/Return-to-Libc attack,
1/Correlated attack,
1/Coordinated attack,
5/Buffer overflow attack,
6/Side-channel attack,
2/Deobfuscation attack,
5/Impersonation attack,
1/Tampering attack,
5/Eavesdropping attack,
3/Denial of Service attack,
4/Reverse engineering}
\legend[shift={(2cm, 1.2cm)}]{{Physical attack}/Physical attack,{Zero-day attack}/Zero-day attack, {Worm attack}/Worm attack, {Code injection attack}/Code injection attack, {Code reuse attack}/Code reuse attack, {Return-to-Libc attack}/Return-to-Libc attack, {Correlated attack}/Correlated attack, {Coordinated attack}/Coordinated attack, {Buffer overflow attack}/Buffer overflow attack, {Side-channel attack}/Side-channel attack, {Deobfuscation attack}/Deobfuscation attack, {Impersonation attack}/Impersonation attack, {Tampering attack}/Tampering attack, {Eavesdropping attack}/Eavesdropping attack, {Denial-of-Service attack}/Denial of Service attack,{Reverse engineering}/Reverse engineering} 
\end{tikzpicture}
\caption{Types and frequency of attacks considered in the existing diversity-based approaches.}
\label{fig:attack_frequency}
\vspace{-3mm}
\end{figure}
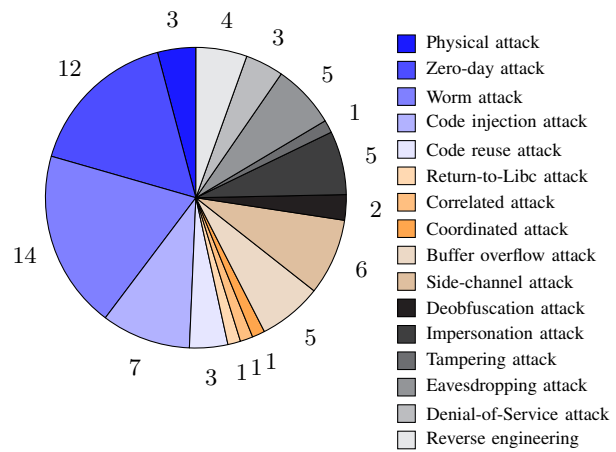

Various types of attacks can be defended by the existing diversity-based approaches.  Due to space constraints, we provided a detailed explanation of each attack in Appendix D of the supplement document.  In this section, to provide the overview of those attacks, we summarized what types of attacks are considered in terms of the number of papers considering each attack type, among the papers discussed in our survey paper in Fig.~\ref{fig:attack_frequency}.  As observed in Fig.~\ref{fig:attack_frequency}, the top three attacks considered in our survey are: worm attack (e.g., malware or virus propagation), zero-day attack, and side channel attack.  Since software diversity is a major trend in diversity-based approaches and software assignment research is mainly studied based on the concept of polyculture software following the fundamental principle of diversity in enhancing system survivability, it is natural to observe more efforts made in mitigating work attacks in the existing approaches.

\section{Metrics, Datasets, and Evaluation Testbeds} \label{sec:v-v}
This section discusses how the existing diversity-based security solutions have been validated by using various types of metrics, datasets, and evaluation testbeds.  Due to space constraints, we provided details of metrics, datasets, and testbeds in the Appendices E-G in the supplement document and discuss the key trends observed from our extensive survey in this section.

\subsection{System Metrics} \label{subsec:metrics}

We discuss the metrics used to validate the existing diversity-based security solutions in terms of measuring security and dependability, respectively.
\subsubsection{\bf Security Metrics} 
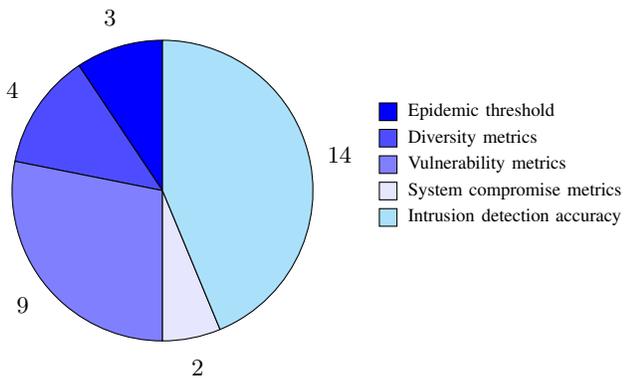
\begin{figure}[h]
\vspace{-3mm}
\centering
\begin{tikzpicture}
[pie chart,
slice type={Epidemic threshold}{blue},
slice type={Diversity metrics}{blue!70},
slice type={Vulnerability metrics}{blue!50},
slice type={Node compromise metrics}{blue!10},
slice type={Intrusion detection accuracy}{cyan!30},
pie values/.style={font={\small}},
scale=2]
\pye[xshift=0.5cm, values of coltello/.style={pos=1}, radius =0.5]{1994}
{
3/Epidemic threshold,
4/Diversity metrics,
9/Vulnerability metrics,
2/Node compromise metrics,
14/Intrusion detection accuracy}
\legend[shift={(2cm, 0.7cm)}]{{Epidemic threshold}/Epidemic threshold, {Diversity metrics}/Diversity metrics, {Vulnerability metrics}/Vulnerability metrics, {System compromise metrics}/Node compromise metrics, {Intrusion detection accuracy}/Intrusion detection accuracy}
\end{tikzpicture}
\caption{Types and frequency of security metrics.}
\label{fig:security-metrics}
\vspace{-3mm}
\end{figure}

Although various types of security metrics have been used to evaluate diversity-based approaches to enhance system security, we identified the following major categories of metrics used in the literature: (i) Epidemic thresholds representing the rate of infecting other nodes in malware or virus propagation; (ii) The extent of diversity measured in code, instructions, or routing paths; (iii) The metrics to capture system vulnerability (or exploitability) to attacks; (iv) The extent of compromised nodes or compromised routes in a given system or network; and (v) Intrusion detection accuracy in diversity-based IDSs. The detailed description of each metric belonging to one of these categories is provided in Appendix E of the supplement document.  Based on the summary of these trends in Fig.~\ref{fig:security-metrics}, the majority of the diversity-based approaches have estimated the system security level based on the system's vulnerability to attacks.  Although it seems clear that high diversity enhances system security, the adverse effect of using a diversity-based approach on system performance has not been thoroughly investigated.

\subsubsection{\bf Dependability Metrics}  As discussed in Section~\ref{subsec:DS-attributes}, dependability embraces availability, reliability, safety, integrity, and maintainability. We extensively surveyed dependability metrics that have been used to validate the quality of diversity-based approaches. However, due to the space constraint, we provided the detail of each dependability metric in Appendix F of the supplement document. Instead, here we discuss the overall trends found from our survey on the dependability metrics in diversity-based approaches.

\begin{figure}[h]
\vspace{-3mm}
\centering
\small 
\begin{tikzpicture}[scale=1]
\pie[sum=auto , after  number=, radius =1.5]
{
7/QoS metrics,
2/Maintenance cost,
4/Reliability}
\end{tikzpicture}
\caption{Types and frequency of dependability metrics.}
\label{fig:dependability-metrics}
\vspace{-5mm}
\end{figure}
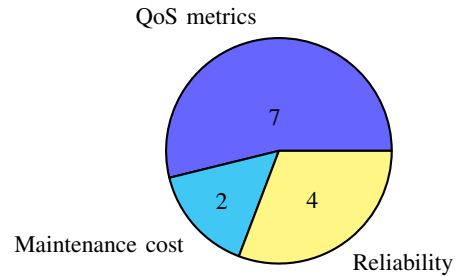

In Fig.~\ref{fig:dependability-metrics}, we summarized the types and frequency of dependability metrics used in the existing diversity-based approaches.  The major trends are: (1) Quality-of-Service (QoS) metrics are the dominant metrics used to capture system dependability, such as packet delivery or loss rates or delay; (2) Reliability is also captured based on load reduction caused by attacks; and (3) Maintenance cost is also observed, such as the financial cost to maintain multiple software packages (or versions).

\subsection{Datasets} \label{subsec:datasets}

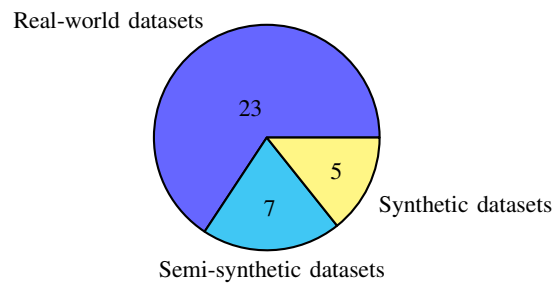
\begin{figure}[h]
\vspace{-3mm}
\centering
\small 
\begin{tikzpicture}[scale=1]
\pie[sum=auto , after  number=, radius =1.5]
{
23/Real-world datasets,
7/Semi-synthetic datasets,
5/Synthetic datasets
}
\end{tikzpicture}
\caption{Types and frequency of datasets used.}
\label{fig:dataset-frequency}
\vspace{-3mm}
\end{figure}

We examined 35 papers that have proposed diversity-based system design for secure and dependable CPSs.  Based on this survey, we could categorize the following three types of datasets used for the validation of the proposed mechanisms: real-world datasets, semi-synthetic datasets, and synthetic datasets. As the names explain, the real-world datasets means the data have been captured from real world environments, such as network traffics or attacks observed in real systems. The synthetic datasets are data generated by simulation that mimic the real world datasets.  Sometimes when researchers cannot find the appropriate dataset to evaluate their proposed mechanism, they combined a real world dataset with synthetic dataset in order to make a dataset that can test the system security and dependability of their proposed mechanism.  Due to space constraints, we provided Table IV in the supplement document that provides the detail of each paper, 35 papers in total.  In Fig.~\ref{fig:dataset-frequency}, we simply summarize the frequency of each dataset type used among the 35 papers surveyed in this work.

Based on Fig.~\ref{fig:dataset-frequency}, we can clearly observe that most studies leveraged real world datasets to evaluate their proposed diversity-based approaches while only 5 works relied solely on synthetic datasets. But based on Table VI of the supplement document, we found the most synthetic datasets are mainly for generating synthetic network topologies.  Although various types of network datasets are available, there is still a limited amount of real `communication network' datasets, resulting in generating synthetic datasets for network topologies. 

\definecolor{software}{HTML}{FFC0CB}
\definecolor{hardware}{HTML}{9BDDFF}
\definecolor{network}{HTML}{FAD6A5}
\definecolor{others}{HTML}{ABDDA4}

\newcommand{\networktrans}[1]{\textcolor{network}{#1}}
\newcommand{\softwaretrans}[1]{\textcolor{software}{#1}}
\newcommand{\hardwaretrans}[1]{\textcolor{hardware}{#1}}

\newcommand{\analytical}{\cite{ghourab2017towards}\cite{ghourab2019spatiotemporal}\cite{sarkar2012enhancing}\cite{zou2015improving}\cite{Gorbenko2019}\cite{gu2008principled}\cite{Hole2013}\cite{Cohen93}\cite{Xu03}\cite{collberg1998manufacturing}\cite{just2004review}\cite{Neti12}\cite{akhuseyinoglu17}\cite{cao20}\cite{clark1987comparison}\cite{shamir1979share} }
\newcommand{\analyticalone}{\cite{clark1987comparison} \phantom{1} \cite{akhuseyinoglu17} \phantom{11} \cite{Neti12} \cite{just2004review} \cite{gu2008principled} \cite{collberg1998manufacturing} \cite{Cohen93} \cite{sarkar2012enhancing} \cite{ghourab2017towards} }
\newcommand{\analyticaltwo}{\cite{shamir1979share} \cite{cao20} \networktrans{\cite{Neti12}}   \cite{Xu03} \cite{Hole2013} \cite{Gorbenko2019} \softwaretrans{\cite{Neti12}} \cite{zou2015improving} \cite{ghourab2019spatiotemporal} }

\newcommand{\simulation}{\cite{karam17}\cite{Calton96}\cite{Qu2018}\cite{gu2008principled}\cite{gondal2015network}\cite{zhang2016network}\cite{temizkan2017software}\cite{borbor2019optimizing}\cite{o2004achieving}\cite{touhiduzzaman2018diversity}\cite{huang2017software}\cite{hong2017optimal} }

\newcommand{\simulationone}{ \cite{touhiduzzaman2018diversity} \cite{o2004achieving} \cite{hong2017optimal} \cite{borbor2019optimizing} \cite{Calton96} \cite{gondal2015network} \cite{karam17} }
\newcommand{\simulationtwo}{ \cite{zhang2016network} \cite{temizkan2017software} \cite{huang2017software} \networktrans{\cite{Qu2018}} \cite{Qu2018} \cite{gu2008principled} }

\newcommand{\emulation}{\cite{Silva2010}\cite{Kc03}\cite{Barrantes03}\cite{hu2006secure}\cite{sovarel2005s}\cite{skandhakumar12} }

\newcommand{\emulationone}{\cite{skandhakumar12}  \cite{sovarel2005s} \cite{hu2006secure} \cite{Barrantes03} }
\newcommand{\emulationtwo}{ \cite{Silva2010} \cite{Kc03} }

\newcommand{\realtestone}{\cite{reiter1996distributing} \cite{Williams09} \cite{Reynolds2002}  \cite{Oberheide2008} \cite{liu2008diverse} \cite{kuang2016exploiting} \cite{ichikawa2008diversification} \cite{hataba2015diversified}  \cite{gashi2009experimental} \cite{Franz10} \cite{cox2006n} \cite{Carvalho2014CloudCA} \cite{Bishop11}  \cite{wang2019diversity} \cite{gerdes2006device} \cite{foruhandeh2019simple} \cite{danev2009transient}}
\newcommand{\realtesttwo}{\phantom{[123]} \cite{xue2018exploiting} \cite{Xu03} \cite{smutz2016tree} \cite{pawlowski2016probfuscation} \cite{majorczyk2007experiments} \cite{kuang2018enhance} \cite{Koo16} \cite{Homescu13} \cite{gashi2009experimental}  \cite{crane2015thwarting} \cite{Chew02} \softwaretrans{\cite{watteyne2009reliability}} \cite{zeng10} \cite{watteyne2009reliability} \cite{lach1999algorithms} \hardwaretrans{\cite{watteyne2009reliability}} }

\newcommand{\realtest}{\cite{danev2009transient}\cite{wang2019diversity}\cite{lach1999algorithms}\cite{zeng10}\cite{foruhandeh2019simple}\cite{gerdes2006device}\cite{watteyne2009reliability}\cite{gashi2009experimental}\cite{liu2008diverse}\cite{Reynolds2002}\cite{cox2006n}\cite{gu2008principled}\cite{majorczyk2007experiments}\cite{Oberheide2008}\cite{Bishop11}\cite{smutz2016tree}\cite{carvalhocloud}\cite{Homescu13}\cite{Koo16}\cite{Chew02}\cite{Williams09}\cite{Franz10}\cite{Xu03}\cite{ichikawa2008diversification}\cite{kuang2016exploiting}\cite{kuang2018enhance}\cite{xue2018exploiting}\cite{pawlowski2016probfuscation}\cite{crane2015thwarting}\cite{hataba2015diversified}\cite{reiter1996distributing} }

\begin{figure}
\centering
\footnotesize
\begin{tikzpicture}[x={(.01,0)}]
\foreach  \l/\x/\y/\z/\h/\s/\o in {Real/100/5.7/31/3.7/1/1.07,Analytical/700/3.15/16/3.5/1.46/1.27,Simulation/500/2.53/12/4.3/2.1/1,Emulation/300/1.6/6/6000/1/1.30} {\node[below] at (\x,0) {\l};
\draw[very thin,fill=hardware] (\x-70,0) rectangle (\x+80,\y/\h);
\draw[very thin,fill=software] (\x-70,\y/\h) rectangle (\x+80,\y/\s);
\draw[very thin,fill=network] (\x-70,\y/\s) rectangle (\x+80,\y/\o);
\draw[very thin,fill=others] (\x-70,\y/\o) rectangle (\x+80,\y);


\node[above] at (\x, \y) {\z};
}

\foreach  \x/\y/\rone/\rtwo in {100/5.7/\realtestone/\realtesttwo,700/3.15/\analyticalone/\analyticaltwo,500/2.53/\simulationone/\simulationtwo,300/1.60/\emulationone/\emulationtwo}{
\fill[fill=none] (\x-80,0) rectangle (\x-10,\y) node[align=left,text width = 1cm, text height = 5.4,anchor=north] {\rone};
}

\foreach  \x/\y/\rone/\rtwo in {100/5.7/\realtestone/\realtesttwo,700/3.15/\analyticalone/\analyticaltwo,500/2.53/\simulationone/\simulationtwo,300/1.28/\emulationone/\emulationtwo}{
\fill[fill=none] (\x-10,0) rectangle (\x+60,\y) node[align=left,text width = 1cm, text height = 5.4,anchor=north] {\rtwo};
}

\foreach  \l/\c/\y in {Hardware/hardware/5, 
Software/software/4.5, 
Network/network/4, 
Others/others/3.5}
{
\draw[fill=\c] (400,\y-.2) rectangle (460,\y+.2) node[below,shift={(60,0)}] {\l};
}

\draw (0,0) -- (800,0);
\foreach \y/\z in {1.25/6,2.8/16,5.3/30}
{\draw (0,\y) -- (5,\y) node[left] {\z};}
\draw (0,0) -- (0,5.8);
\end{tikzpicture}
\caption{Evaluation Testbeds Used for Existing Diversity-based Approaches.}
\label{fig::testbed}
\vspace{-3mm}
\end{figure}
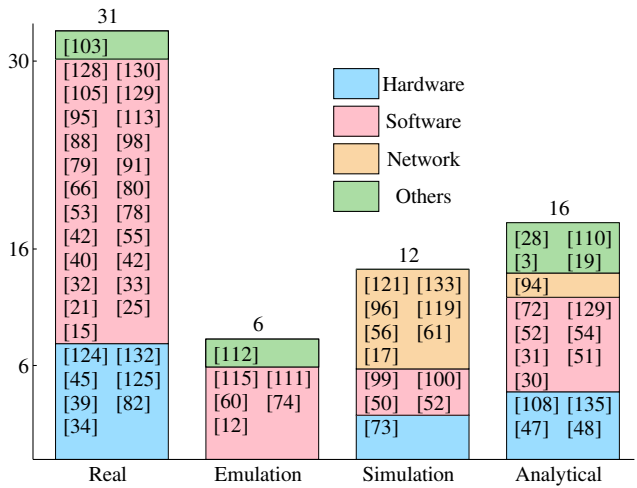

\subsection{Evaluation Testbeds} \label{subsec:testbeds}

In the literature, we found mainly the following four types of evaluation testbeds being used: real testbeds, emulation-based testbeds, simulation testbeds, and analytical or probability model-based testbeds.  In Section~\ref{sec:diversity-cybersecrity-techniques}, we also discussed the multiple layers a diversity-based solution is deployed at.  To grasp the overall picture of what testbeds have been used to validate the diversity-based approaches that are developed at a certain layer, we summarized the testbeds used for the validation of the existing diversity-based approaches in Fig.~\ref{fig::testbed}. In Fig.~\ref{fig::testbed}, each color represents the layer at which a diversity-based approach is deployed while we showed what evaluation testbed is used to validate each approach.  Interestingly, real testbeds are popular for developing both hardware and software diversity-based approaches.  The second most popular validation testbed was the analytical model-based one.  Particularly for network diversity-based approaches, simulation-based testbeds are popularly used in the existing works.

These observations are well aligned with the datasets used to validate the existing approaches. As discussed in Section~\ref{subsec:datasets}, real-world datasets are the most popular in use while synthetic datasets based on simulation models are mainly used for network topologies, which are used in validating network diversity-based approaches.


\section{Limitations, Insights, and Lessons Learned} \label{sec:discussions}
We found the following limitations and learned insights and lessons from this extensive survey:
\begin{itemize}
\item {\bf Fewer studies in identifying critical tradeoffs between system diversity and other aspects of system quality:}  Although it is well-known that diversity-based system designs can significantly enhance system security, it may not be always true~\cite{Chen18-network-diversity}. For example, if each software has inherently high vulnerabilities, increasing diversity with poor software components would not contribute to increasing system security. For example, using a set of diverse detectors may not necessarily lead to high system security. When each detector's detection capability is too poor (e.g., $< 0.5$), the system is still highly vulnerable due to the misdetection by the poor detectors. Although diversity-based design can be easily integrated with legacy security mechanisms and existing technologies, there should be studies investigating the critical tradeoff setting to identify the bottom line for achieving the benefit of diversity-based system designs.

\item {\bf Lack of research investigating the drawbacks of diversity-based system designs}:  It is well-known that introducing more diversity to the system can bring adverse impact on configuration cost, service availability, and economic cost.
 However, there has been less effort in investigating the key drawbacks of diversity-based system designs and how to mitigate the drawbacks.

\item {\bf Need more effort to explore diversity-based system designs in broader areas:} Based on our extensive survey in diversity-based system designs, we found that OSs, IDSs, malware detection, and instruction diversifications have been substantially studied. However, there have been significantly fewer studies developing diversity-based security mechanisms in firewalls and cryptographic authentication.  The reason would probably be the benefit of using diversity-based approaches would not exceed that of not using them.  However, no clear investigation has been even conducted to answer this.  Based on the critical tradeoff analysis of using diversity and not using it, we can set our research towards a more promising direction.

\item {\bf Less adverse impact of diversity-based designs at lower layers on system performance:}  Diversity-based designs deployed at a lower layer (i.e., instruction diversification) tend to have less adverse impact on system performance.  On the other hand, when diversity-based designs are considered at higher layers, computational or memory resources tend to be more often required.

\item {\bf Integration of hardware diversity and software diversity}: Although there has been a fairly good amount of diversity-based approaches by introducing software diversity or hardware diversity, we have not found any research effort to explore diversity of both hardware and software and investigate the impact of the integrated approaches.

\item {\bf Lack of research examining the relationships between diversity and other system dependability and security attributes:} As we can observe from Table~\ref{tab:diversity_evolution}, until the 2000's, the primary effort of diversity-based system design was to increase software fault tolerance. Even if there have been more studies explored in the 2000's and the 2010's for investigating system security and dependability attributes, there have been many works that are still focused on enhancing fault tolerance. In addition, the relationships between diversity and other system attributes, such as confidentiality, maintainability, safety, and so forth, are still unclear.  

\item {\bf Lack of deploying diversity-based approaches under dynamic system environments:} Some recent efforts have explored diversity-based approaches under dynamic system environments~\cite{Cho19}. However, most current diversity-based research has been studied under static system environments where system components are fixed once diversity is implemented, such as code diversification, malware detectors, code instruction diversification or obfuscation, and so forth. 

\item {\bf Limited theoretical understanding of diversity-based approaches:} Most diversity-based approaches have been validated based on simulation or emulation testbeds.  Surely, the extensive experiments via simulation and emulation can provide a certain level of confidence on proposed technologies. However, validating their effectiveness and efficiency via mathematical and analytical models can further provide a solid basis of demonstrating their powerful merits on system security and dependability.

\item {\bf Lack of valid diversity metrics}: Most literature surveyed in our paper have not devised or used diversity metrics to quantify system diversity. Even though there are some studies that have proposed diversity metrics and their comparative analysis particularly in software assignment research~\cite{borbor2019optimizing, lyu1992software, temizkan2017software}, there is still a lack of studies that conduct in-depth analysis of various types of diversity metrics.
\end{itemize}

\section{Conclusion and Future Work} \label{sec:conclusion}
In this section, we conclude this work by summarizing the key findings obtained from this extensive survey. And then, we suggest future work directions to develop diversity-based solutions to build secure and dependable CPSs.

\subsection{Key Findings}
From our extensive survey on diversity-based approaches, we obtained the following key findings:
\begin{itemize}
\item The key principle of diversity-based system designs is to enhance resilience, survivability, or sustainability of the system by increasing attack cost or complexity for attackers to compromise the system by exploiting the same system vulnerabilities. However, deploying diversity-based approaches may introduce additional cost or performance degradation due potential cross incompatibility issues, maintenance cost, or high dynamic system/network reconfigurations.  As a diversity-based solution designer, we should consider critical tradeoffs that can optimize the effectiveness and efficiency of diversity-based mechanisms.

\item While software diversity, hardware diversity, and network diversity are the three most popular approaches used in the literature, diversity-based solutions for physical environments and human-machine interactions to enhance security and dependability are rarely explored.

\item Although diversity-based approaches for enhancing system security has been explored since the 1970's and both system security and dependability since the 2000's, the maturity of diversity metrics has not been reached for them to be used as general metrics like other security or dependability metrics (e.g., mean time to security failure, reliability, or availability). Entropy has been commonly used to capture uncertainty, representing a measure of randomness where higher diversity is assumed to show high uncertainty. However, as high entropy can be shown when there are not many variants of system components, it is highly questionable to simply use entropy as a diversity metric.

\item We found that the three most popular attacks considered in the existing diversity-based approaches are worm attacks, zero-day attacks, and code injection attacks based on our survey. Since software diversity-based approaches are popularly used to increase network diversity, it is natural to observe that worm attacks performing epidemic attacks (e.g., malware or virus propagation) are the most popular attack type considered in the existing diversity-based techniques. 
\item We found that most diversity-based approaches use existing security metrics to capture their effect on security although it is not crystal clear that diversity can enhance security regardless of context or environmental conditions. Most security metrics are mainly based on the extent of system vulnerability to cyberattacks. The existing diversity-based approaches have also used dependability metrics that are most often used to measure Quality-of-Service (QoS) metrics (e.g., message delivery ratio, throughput, delay) while pure dependability metrics, including availability, reliability, or performability, have not been sufficiently considered. 

\item Unlike other cybersecurity research domains, in diversity-based research, many studies used real datasets and real testbeds to validate the proposed diversity-based approaches (see Figs.~\ref{fig:dataset-frequency} and \ref{fig::testbed}). Most synthetic datasets and simulation models are used to evaluate network diversity-based approaches where the datasets represent network topologies and simulation models are used to evaluate network resilience under various epidemic attacks.  
\end{itemize}

\subsection{Future Work Directions}

According to the lessons learned from Section~\ref{sec:discussions}, we suggest the following future research directions to develop diversity-based approaches to enhance security and dependability of CPSs:
\begin{itemize}
\item {\bf Investigate critical tradeoffs between system diversity and other aspects of metric attributes}.  For example, we need to clarify the relationships between diversity and security where each system component's vulnerabilities vary. In addition, we should examine what other drawbacks can be introduced by using diversity-based approaches, such as performance degradation, maintenance cost, cross-incompatibility, and so forth.

\item {\bf Broaden the areas to deploy/apply diversity-based approaches}. We rarely found existing diversity-based approaches in certain areas, such as firewalls, cryptographic authentication, physical environments, and/or human-machine interactions.  In these areas,  we should investigate if diversity-based approaches can introduce more benefits based on their key advantages.


\item {\bf Integrate hardware and software diversity-based approaches}. Although substantial efforts are made in both software and hardware diversity research, there has been no research integrating both. Investigating the feasibility and merits of combining them should be the first step to start this research.

\item {\bf Develop dynamic diversity-based approaches to enhance system security and performance}. Environmental conditions and their dynamics require vastly different approaches to tackle the development of diversity-based approaches. We should take a first step to tackle this problem in terms of what additional overhead can be introduced to deploy dynamic diversity-based approaches while how much security and dependability can be enhanced even under the dynamic contexts by leveraging diversity-based solutions.

\item {\bf Develop meaningful metrics including diversity metrics}.  As discussed in Section~\ref{subsec:metrics}, the current research in diversity-based system designs mostly uses existing metrics that cannot capture the clear merit of diversity-based approaches. In addition, as the platforms of the diversity-based approaches being deployed become more diverse, there should be more relevant metrics to capture diversity and its effect on system security and dependability.

\item {\bf Explore the theoretical validation of diversity-based approaches}.  As shown in Figs.~\ref{fig::testbed} and \ref{fig:dataset-frequency}, unlike other research domains, we observed real-world datasets and real testbeds have been popularly used in the existing diversity-based approaches.  However, theoretical validation was relatively weakly observed.  It is fundamental to provide the theoretical basis of an approach because the theoretical validation can provide a generic framework that can enable other researchers to easily adopt a given approach.
\end{itemize}

\bibliographystyle{IEEEtranSN}
\bibliography{ref}

\end{document}


\pagenumbering{arabic}

\title{Diversity-By-Design for Dependable and Secure Cyber-Physical Systems: A Survey}
\author{Qisheng Zhang, Abdullah Zubair Mohammed, Zelin Wan, Jin-Hee Cho, \IEEEmembership{Senior Member, IEEE}, and Terrence J. Moore, \IEEEmembership{Member, IEEE}\IEEEcompsocitemizethanks{\IEEEcompsocthanksitem Qisheng Zhang, Zelin Wan, and Jin-Hee Cho are with The Department of Computer Science, Virginia Tech, Falls Church, VA, USA. Email: \{qishengz19, zelin, jicho\}@vt.edu.  Abdullah Zubair Mohammed is with The Bradley  Department of Electrical and Computer Engineering, Virginia Tech, Arlington, VA, USA.  Email: abdullahzubair@vt.edu.  Terrence J. Moore is with US Army Research Laboratory, Adelphi, MD, USA. Email: terrence.j.moore.civ@mail.mil.  The first three authors made almost a same amount of contributions.}}

\maketitle

\pagenumbering{arabic}

\appendices

\section{Comparison between Our Survey Paper and Existing Survey Papers}

We found several similar survey papers~\cite{Balakrishnan05, Baudry15-survey, hosseinzadeh2016, Hosseinzadeh18-survey, Larsen14-survey} to our survey paper in the literature. Here we discuss each work to clarify the key contributions of our survey paper compared to these existing works.

\citet{Balakrishnan05} conducted a limited survey based on 14 references and mainly focused on code obfuscation, a typical method to apply software diversity at an instruction level.  They covered general code obfuscation techniques used by viruses to hide malicious activities or commercial software to protect valuable assets.  Instead of directly showing applications of code obfuscation in the protection of assets, this paper analyzed the attack patterns of a virus to show the potential usage of code obfuscation in the security domain.  However, this paper didn't provide any security or performance metrics used in the existing obfuscation techniques. 

\citet{Baudry15-survey} surveyed recent studies on software diversity. They categorized each work into either managed diversity or automated diversity based on the human involvement. They discussed the core concept of each technique along with its pros and cons.  As for performance evaluations, the authors primarily discussed resilience and security to show the effectiveness of diversity techniques. They also provided an overview on the research history of software diversity. However, their work didn't conduct a comprehensive survey on attacks covered and performance metrics used by the existing diversity-based security techniques, as we did in our survey paper.

\citet{Larsen14-survey} conducted a more extensive survey on software diversity than~\cite{Balakrishnan05, Baudry15-survey}.  Their survey paper provided two different types of classification of software diversity: (i) the scope and level where diversifications occur, including instruction, basic block, loop, function, program, and system levels; and (ii) the development time when diversifications occur, such as implementation, compilation, linking, installation, loading, executing, and updating.  They also discussed extensive sets of attacks and diversity metrics employed in the existing diversity-based security approaches.  For each work cited in their survey paper, the authors discussed the performance of a corresponding diversity technique evaluated by running time and memory.  They discussed the pros and cons of each diversity technique, providing an example.

\citet{Hosseinzadeh18-survey} reviewed over three hundred papers related to diversification and obfuscation from 1993 to 2017. This survey provided an overview of the current existing research in this area. Specifically, they categorized papers based on their aims, attacks mitigated, environments used and development time as in~\cite{Larsen14-survey}. However, this work mainly introduced different software diversity techniques other than analyzing their pros and cons in detail.  The authors also provided the summary of evaluation methods in performance (e.g., memory size, running time) and security against attacks.

\begin{table*}[ht]
\centering
\caption{Comparison of Our Survey and Other Existing Surveys on Diversity-based System Designs}
\label{tab:other_survey}
\vspace{-3mm}
\begin{tabular}{|P{6.5cm}|P{1.4cm}|P{1.4cm}|P{1.4cm}|P{1.4cm}|P{1.4cm}|P{1.4cm}|}
\hline
Criteria & Our survey paper & \citet{Balakrishnan05} (2005) & \citet{Larsen14-survey} (2014) &  \citet{Baudry15-survey} (2015) &
\citet{hosseinzadeh2016} (2016)&
\citet{Hosseinzadeh18-survey} (2018) \\
\hline
\hline
Multidisciplinary concept of diversity & \checkmark & \xmark & \xmark & \xmark & \xmark
&\xmark\\
\hline
Design principles: benefits \& caveats & \checkmark & \checkmark & \checkmark &  \checkmark &\xmark
&Limited (summary) \\
\hline
Attributes \& properties of dependable and secure CPSs  & \checkmark & \xmark& \xmark & \xmark & \xmark&
\xmark\\
\hline
Diversity-based approaches discussed based on multi-faceted dimensions of CPSs  & \checkmark & Limited (obfuscation techniques)& \checkmark & \checkmark& Limited (summary) & \checkmark\\
\hline
Attacks countermeasured by diversity-based approaches & \checkmark & \xmark & \checkmark & \xmark& Limited (Statistical table) & Limited (Statistical table)\\
\hline
Validation \& verification methodologies (i.e., metrics, datasets, and evaluation testbeds) & \checkmark & \xmark & \xmark &Limited (cost) & \xmark
&Limited (cost) \\
\hline
Discussions on insights, limitations, and lessons learned & \checkmark & \checkmark& \checkmark& \checkmark&\checkmark &
\checkmark \\
\hline
\end{tabular}
\end{table*}

\section{Multidisciplinary Concepts of Diversity}

In this section, we survey the concepts of diversity in 9 different disciplines, including, biodiversity, geodiversity, biology, sociology, psychology, political science, organizational management, nutrition science, and computing \& engineering. In particular, based on this comprehensive survey on the concept of diversity, we discuss how the multidisciplinary concept of diversity has been applied to build dependable and secure CPSs.

\subsubsection{\bf Biodiversity}
\citet{DeLong96} defined biodiversity as a property of a region that deals with the variety between or within living organisms, communities and processes that occur naturally or are introduced by humans.  Biodiversity offers a multi-functionality of services in the ecosystem, including the stability of habitats, the regulation of climate, water and gas, air purification, erosion control, nutrient cycling, among others~\cite{Turner07-bioscience}. \citet{Kennedy02} also discussed how biodiversity can enhance the {\em resilience} of an ecosystem by mitigating foreign invasions with rich kinds of species.

\subsubsection{\bf Geodiversity}
Geodiversity refers to the heterogeneity among the abiotic components and processes of the earth, including geological (e.g., rocks and minerals), geomorphological (i.e., the processes of the earth's surface, soil, and water~\cite{Brilha18-env-science}.  \citet{Schrodt19} emphasized the importance of geodiversity in securing our ecosystem, in addition to biodiversity, with the goal of establishing sustainable ecosystems. 

\subsubsection{\bf Biology}
\label{multidisc_subsec_bio}
Genetic variation among each individual in a species introduces diversity in biological systems of living organisms~\cite{ Kephart99-immune-com-sys, Somayaji97}.  The immune system, such as in vertebrates, draws the best analogy with a computer network. A type of white blood cell, called {\em lymphocyte}, is an essential part of the immune system fighting against disease-causing agents, called {\em pathogens}. Pathogens have a toxic substance called an {\em antigen}, which triggers the lymphocyte to generate antibodies to fight against the pathogen~\cite{Somayaji97}.  Different lymphocytes react to different antigens. Therefore, to cover all the antigens, the immune system has a diverse set of lymphocytes. Of course, no single individual's immune system is complete. A pathogen that does not affect one individual may affect another one. However, as a species, not all individuals are vulnerable to specific pathogens. This implies that some individuals may be affected by a particular pathogen while it does not affect the whole population and therefore helps the survival and sustainability of the species.  Biology-inspired computer systems based on the philosophy of the immune system have received significant interest in the engineering fields~\cite{Farah18-immune, Harmer02-immune, Kephart99-immune-com-sys, Roman19, Somayaji97}.

\subsubsection{\bf Sociology} The sociological perspective of diversity research spans the investigation of differences in demographic traits (e.g., race, ethnicity, gender, age, marital status), sexuality, socio-economic status, religion, nationality, appearance, ability, and experience~\cite{Clark99}. Sociologists have discussed the benefit of diversity as a way to facilitate social sustainability~\cite{Caulfield01-sociology}.

\subsubsection{\bf Psychology}
Environmental psychologists have studied ``the concept of affinity towards diversity (ATD)'' that can psychologically drive `pro-sustainability orientation' and `environmentally friendly behaviors'~\cite{Corral-Verdugo09}.  In addition, they also emphasized the importance of diversity as a driving force for sustainability that provides solidarity as the capability of cooperation and mutual aids~\cite{Jimenez-Dominguez02}. Viewpoints diversity is also examined by psychologists. A recent study~\cite{Jose15} looked at evidence showing how political diversity and disagreement can improve the extent of reliability and validity of social science by detecting fallacies and errors in the process of experiments.

\subsubsection{\bf Political Science}
Political scientists mainly studied how political diversity affects of political integration~\cite{Klein05}. Even if people vote for the same candidate in an election, their views are diverse, implying that the political ideology is not necessarily identical. Hence, these studies are predominantly concerned with how diversity in race, ethnic groups, gender, and sexual orientation can affect political propensity~\cite{Crouch06}.  Political scientists~\cite{Hong04} developed a mathematical model to prove how much a diverse group (e.g., with diverse solvers) can outperform in a problem solving exercise requiring creativity compared to the solving ability of a homogeneous group (i.e., with same high-skilled solvers).

\subsubsection{\bf Organizational Management} In this domain, diversity has been studied as a key driver to improve productivity in the workplace~\cite{van04}.  The work diversity is tested in terms of social categorization, information, and decision making skills.  Although homogeneous groups perform better in social categorization, heterogeneous groups performed better in the tasks requiring information and decision making skills.  This finding is also well aligned with that in \cite{West09}, showing higher accuracy in detecting bankruptcy when groups with diverse skills are involved in the decision making process.

\subsubsection{\bf Nutrition}  In this domain, dietary diversity is defined as the different types food consumed by a household in a fixed reference time frame. Several studies have supported the finding that dietary diversity positively impacts food security and nutrition~\cite{Hoddinott02, Steyn06}

\subsubsection{\bf Computing \& Engineering} In this domain, diversity is adopted as a design principle in order to support system dependability, security, and resilience (or fault-tolerance) by utilizing different types of software, hardware, or protocols that provide the same functionalities~\cite{yang2016improving}.  Initially, system diversity has emerged because different vendors produce different hardware or software, which can naturally generate artificial diversity.  But it turned out using the artificial diversity can provide high resistance against cyberattacks because systems with high diversity designs cannot be easily compromised by attackers or even increase attack complexity and cost due to the polyculture software/hardware nature~\cite{Cho19}.  Nature-inspired computing~\cite{Marrow00-nature-inspired} is inspired by living systems that can prolong with high diversity.

\begin{table*}[ht]
\centering
\caption{Multidisciplinary Concepts of Diversity: Key Definitions, Application Domains, Supporting System Quality, and Caveats}
\label{tab:multi-diversity-concepts}
\vspace{-3mm}
 \begin{tabular}{|P{2cm}|P{2.5cm}|P{3cm}|P{3cm}|P{3cm}|P{1.5cm}|} 
 \hline
 {\bf Discipline} & {\bf Key Concept} & {\bf Application Domain} & {\bf Supporting System Quality} & {\bf Caveats} & {\bf Ref.} \\
 \hline\hline
 {\bf Biodiversity} & Diversity in Living organisms and processes & Ecosystems & Sustainability, resilience (or resistance or fault tolerance), stability & Socio-economic costs, impact on other ecosystem services &   \cite{DeLong96, Kennedy02, Turner07-bioscience} \\ 
 \hline
 {\bf Geodiversity} & Diversity in non-living organisms and processes & Ecosystems & Climate regulation, clean energy, sustainability & High cost for diversification & \cite{Brilha18-env-science, Schrodt19} \\
 \hline
 {\bf Biology} & Genetic variations  & Immune system; pharmacology & Disease control, survivability of species & High cost in health care; patients' privacy issue
 & \cite{Farah18-immune, Harmer02-immune, Kephart99-immune-com-sys, Roman19, Somayaji97} \\
 \hline
 {\bf Sociology} & Diversity in demographics or individual differences & The extent of acceptance to society, community, or groups & Social sustainability & Lack of understanding in human diversity & \cite{Caulfield01-sociology, Clark99}\\
 \hline
 {\bf Psychology} & Diverse viewpoints or attitudes towards political leaning, cooperation, or mutual aids & Organizations, communities, society as cooperative entities & Reliability; validity, solidarity, sustainability & Extreme polarization upon failure of accepting diversity & \cite{Corral-Verdugo09, Jimenez-Dominguez02, Jose15}\\
 \hline
 {\bf Political Science} & Diversity in political leaning and/or groups & Election campaign; decision making & Political integration; problem solving skills; sustainability & Risk in low consensus/non-integration & \cite{Crouch06, Hong04, Klein05}\\
  \hline
 {\bf Nutrition} & Health & Human nutrition & Food security; nutrition quality' survivability & Food cost & \cite{Hoddinott02, Steyn06}\\
 \hline
 {\bf Organizational Management} & People in their cultural identities, demographic characteristics, skills, and expertise & Organizations; communities &  Productivity; creativity; problem solving; decision making & Potential delay in converging diverse opinions; lack of understanding in those with other background & \cite{van04, West09} \\
\hline
 {\bf Computing / Engineering} & Natural/Artificial design of computer-based systems & Computer-based systems/networks; software engineering & Fault tolerance; resilience; sustainability; survivability; dependability; security & Cost for diversity-based system designs; interoperability & \cite{Cho19, Marrow00-nature-inspired, yang2016improving} \\ 
 \hline
 \end{tabular}
\end{table*}

\section{Diversity-based Approaches for Secure and Dependable Cyber-Physical Systems}

In Table~\ref{tab:diversity-summary}, we summarized the overview of the existing diversity-based approaches surveyed in Section IV of the main paper. 

\begin{table*}[ht]
\centering
\caption{Techniques of Diversity-based Designs for Secure and Dependable CPSs}
\label{tab:diversity-summary} 
\vspace{-2mm}
\begin{tabular}{|P{0.8cm}|P{3cm}| P{3cm}|P{3cm}|P{3cm}|P{3cm}|} 
\hline
{\bf Ref.} & {\bf Technique Type} & {\bf Attacks Mitigated} & {\bf Metric for Evaluation} & {\bf Pros} & {\bf Cons} \\ [0.5ex] 
\hline
\hline
\multicolumn{6}{|c|}{\cb {\bf Diversity for Physical Environments}} \\
\hline
\hline
\cite{skandhakumar12} (2012) & Geographical topology diversity & Physical attack & No experiments conducted & Provide security solutions for direct physical access & Hard to implement or update\\ 
\cline{1-4}
\cite{akhuseyinoglu17} (2017) & Action sequence diversity & Physical attack & No experiments conducted &  Easy to manage access control systems through graphical environments, such as GPS in 3D maps & Sensitive to human errors \\ 
\cline{1-4}
\cite{cao20} (2020)& Access constraint diversity & Physical attack &  Percentage of granting requests &  &  \\ 
\hline
\hline
\multicolumn{6}{|c|}{\cb {\bf Network Diversity}} \\
\hline
\hline
\cite{temizkan2017software} (2017)& Metric-based software allocation & Worm attack &  Epidemic Threshold & Easy formalization as optimizing problems & High complexity\\ 
\cline{1-4}
\cite{borbor2019optimizing} (2019)& Metric-based software allocation & Zero-day attack& $d_1,d_2,d_3$ metrics & & \\ 
\cline{1-6}
\cite{o2004achieving}  (2004)& Metric-free software allocation & Worm attack & Defective edge count  & Strong theoretic support & Only validated in static networks\\ 
\cline{1-4}
\cite{huang2017software} (2017)& Metric-free software allocation & Worm attack & Maximum size of common vulnerability graphs & High efficiency & Reconfiguration cost \\ 
\cline{1-4}
\cite{touhiduzzaman2018diversity} (2018)& Metric-free software allocation & Coordinated attack & Node exploitability index & & \\
\cline{1-6}
\cite{hong2017optimal} (2017)& Network topology shuffling& Correlated attack & Expected path variant (EPV)& Low implementation cost & Only applicable in dynamic networks\\ 
\hline
\hline
\multicolumn{6}{|c|}{\cb {\bf Hardware Diversity}} \\
\hline
\hline
\cite{lach1999algorithms} (1999) & FPGA architectural diversity & Zero-day attack & Fault-free probability & Flexibility &  Area overhead \\ 
\cline{1-4}
\cite{gerdes2006device} (2006) & Physical Layer Identification and Device Fingerprinting & Impersonation Attack & Intrusion detection accuracy &  Physical Authentication & Training per hardware \\ 
\cline{1-4}
\cite{danev2009transient} (2009) & Physical Layer Identification and Device Fingerprinting & Denial of Service Attack & Intrusion detection accuracy & Intrusion Detection & Sensitive to environment \\ 
\cline{1-4}
\cite{cobb2010physical} (2010) & Physical Layer Identification and RF-DNA Fingerprinting & Impersonation Attack & Average correct classification percentage and Confusion Matrix & Device Identification & Sensitive to aging of hardware\\ 
\cline{1-4}
\cite{foruhandeh2019simple} (2019) & Physical Layer Identification and Device Fingerprinting & Impersonation Attack & Intrusion detection accuracy &  & \\ 
\cline{1-6}
\cite{karam17} (2017) & FPGA architectural diversity & Side-channel attacks, Tampering attack & Inter and intra bit stream distances & Ease of implementation & Latency overhead\\ 
\cline{1-6}
\cite{watteyne2009reliability} (2009) & Frequency Diversity & Zero-day attack & Probability of connection, Average Expected Transmission Count (ETX) and Network churn & Secure communication between nodes & Higher power consumption \\ \cline{1-4}
\cite{zeng10} (2010) & Antenna Diversity & Eavesdropping & Randomness, bit-agreement ratio and shared bit-generation rate & Improvement of network utilization & Increase in hardware complexity \\
\cline{1-4}
\cite{zou2015improving} (2015) & Antenna and User Diversity & Eavesdropping & Secrecy capacity and intercept probability & Improvement of network connectivity 
& \\
\cline{1-4}
\cite{ghourab2017towards} (2017) & Antenna Diversity & Eavesdropping & Secrecy capacity and intercept probability &  &  \\ 
\hline
\end{tabular}
\end{table*}

\begin{table*}[h]
\centering
\vspace{-5mm}
\caption*{ (continued) Techniques of Diversity-based Designs for Secure and Dependable CPSs}
\vspace{-3mm}
\begin{tabular}{|P{1.5cm}|P{2.5cm}| P{2.5cm}|P{3cm}|P{3cm}|P{3cm}|} 
\hline
{\bf Ref.} & {\bf Technique Type} & {\bf Attacks Mitigated} & {\bf Metric for Evaluation} & {\bf Pros} & {\bf Cons} \\ [0.5ex] 
\hline
\hline
\multicolumn{6}{|c|}{\cb {\bf Software Diversity}} \\
\hline
\hline
\cite{Calton96} (1996)& OS diversity & Worm attack & Survivability rates & Dynamic adaptation against attacks & Code complexity overhead\\
\cline{1-6}
\cite{forrest1997building} (1997)& OS diversity & Buffer overflow attack & Stack usage & Enhance robustness with minimal efficiency impact & May disrupt code legitimacy\\ 
\cline{1-6}
\cite{nagy2006n} (2006)& OS diversity & Zero-day attack & Intrusion detection accuracy & Effective against unknown attacks & Labor overhead \\
\cline{1-6}
\cite{Garcia2014} (2014)& OS diversity & Zero-day attack & Common vulnerabilities & Theoretic analysis based on public datasets & No simulations\\ 
\cline{1-4}
\cite{Gorbenko2019} (2019)& OS diversity & Zero-day attack & Common vulnerabilities & & \\
\cline{1-6}
\cite{liu2008diverse} (2008)& Firewall diversity & Zero-day attack & Functional discrepancies & Effective against unknown attacks &  Labor overhead\\
\cline{1-6}
\cite{Reynolds2002} (2002) & Intrusion Detection & Worm attack& Failover on attack & Detection of threats & False alarms  \\
\cline{1-4}
\cite{Totel2005} (2005) & Intrusion Detection & Denial of Service attack & Intrusion detection accuracy & Increase of detection coverage against many attacks by using diverse IDS  & Lack of proactive defense (e.g., prevention) \\
\cline{1-4}
\cite{cox2006n} (2006) & Intrusion Detection & Code injection attack & Intrusion detection accuracy & Enhancement of tracking the attack pattern & Performance and maintenance cost \\
\cline{1-4}
\cite{majorczyk2007experiments} (2007) & Intrusion Detection & Generic intrusions & Intrusion detection accuracy &  & IDS being attacked by DoS attacks to create false alerts \\
\cline{1-4}
\cite{gu2008principled} (2008) & Intrusion Detection & HTTP and TCP intrusions & Intrusion detection accuracy & & \\
\cline{1-4}
\cite{gondal2015network} (2015) & Intrusion Detection & Denial of Service attack & Intrusion detection accuracy & & \\
\cline{1-4}
\cite{Qu2018} (2018) & Intrusion Detection & Code injection attack & Intrusion detection accuracy & & \\
\cline{1-6}
\cite{gashi2009experimental} (2009) & Antivirus diversity & Worm attack & Intrusion detection accuracy & Mitigation of zero-day attacks & High computing resources\\
\cline{1-6}
\cite{Carvalho2014CloudCA} (2014) & Cryptographic algorithm diversity & XML signature warping attack & Algorithm specific metrics & Multifold security& Performance and complexity overhead \\
\hline
\cite{collberg1998manufacturing} (1998) & Obfuscation & Reverse engineering & Measures of potent, resilience, stealth and cost  &  Increases the cost for an attacker.  & Cannot guarantee the irreversibility of code.   \\
\cline{1-4}
\cite{just2004review} (2004) & Obfuscation & Worm attack & Attack specific & Increase the time required for an attacker. & Techniques using  VMs in parallel requires tremendous computational resources.\\
\cline{1-4}
\cite{hataba2015diversified} (2015)& Obfuscation & Side-channel attack & Code similarity &  Increase of trace analysis complexity & \\
\cline{1-4}
\cite{crane2015thwarting} (2015)& Obfuscation & Side-channel attack & Number of key-bits discovered by attacker & High challenge in reverse engineering & \\
\cline{1-4}
\cite{pawlowski2016probfuscation} (2016) & Obfuscation & Dynamic trace analysis & Code similarity & & \\
\cline{1-4}
\cite{kuang2016exploiting} (2016) & Obfuscation & Reverse engineering & Calling frequency & & \\
\cline{1-4}
\cite{xue2018exploiting} (2018) & Obfuscation & Reverse engineering & Likelihood of mapping & & \\
\hline

\end{tabular}
\end{table*}

\begin{table*}[ht]
\centering
\caption*{(continued) Techniques of Diversity-based Designs for Secure and Dependable CPSs}
\vspace{-2mm}
\begin{tabular}{|P{0.8cm}|P{3cm}| P{3cm}|P{3cm}|P{3cm}|P{3cm}|} 
\hline
{\bf Ref.} & {\bf Technique Type} & {\bf Attacks Mitigated} & {\bf Metric for Evaluation} & {\bf Pros} & {\bf Cons} \\ [0.5ex] 
\hline
\hline
\multicolumn{6}{|c|}{\cb {\bf Software Diversity} (continued)} \\
\hline
\hline
\cite{Chew02} (2002) & Code diversity: OS randomization & Buffer overflow attack & None  & Retaining functionality & Storage overhead \\
\cline{1-4}
\cite{Xu03} (2003) & Code diversity: TRR & Code injection attack & Logical explanation  &  Increase in difficulty for am attacker in predicting memory addresses &  Performance overhead\\
\cline{1-4}
\cite{Barrantes03} (2003) & Code diversity: ISR & Code injection attack & Intrusion detection accuracy  & Increase of attack cost or complexity in forming gadgets & Lack of estimating ASLR addresses using side-channel attacks \\
\cline{1-4}
\cite{Kc03} (2003) & Code diversity: ISR & Code injection attack & Intrusion detection accuracy  & Increased difficulty in reverse engineering & High likelihood of machine level ISR techniques vulnerable to incremental key-breaking attacks \\
\cline{1-4}
\cite{hu2006secure} (2006) & Code diversity: ISR using AES  & Code injection attack & Intrusion detection accuracy  & Increase of the heterogeneity of a system & \\
\cline{1-4}
\cite{ichikawa2008diversification} (2008) & Code diversity: ISR & Plagiarism, analysis & Attack specific  & Protection against a wide variety of buffer overflow attacks &  \\
\cline{1-4}
\cite{Williams09} (2009) & Code diversity: ISR and Calling sequence & Code injection attack & Manual evaluation  & & \\
\cline{1-4}
\cite{Homescu13} (2013) & Code diversity: NOP insertion & Code reuse attack  & Survivability rates & & \\
\cline{1-4}
\cite{Koo16} (2016) & Code diversity: Instruction displacement & Code reuse attack & Randomization code coverage  & & \\ \cline{1-4}
\cite{Franz10} (2010) & Compiler diversity & Reverse engineering & None  & & \\
\hline
\hline
\multicolumn{6}{|c|}{\cb {\bf Diversity for Human-Machine Interactions}} \\
\hline
\hline
\cite{clark1987comparison} (1987) & Human operator diversity & Zero-day attack & No experiments conducted & Ensure the external data consistency & Labor overhead\\ 
\cline{1-4}
\cite{reiter1996distributing} (1996) & Human operator diversity & Zero-day attack & No experiments conducted & & \\ 
\cline{1-6}
\cite{deswarte1998diversity} (1998) & Human error diversity & Zero-day attack & No experiments conducted & Introduce inherent software design diversity & Error overhead\\ 
\cline{1-4}
\cite{huang2014links} (2014) & Human error diversity & Zero-day attack & No experiments conducted & & \\ 
\hline
\end{tabular}
\end{table*}

\section{Attack Types Considered by Diversity-based Security Approaches}

The existing diversity-based security techniques have been designed and used to counter the following attacks: 
\begin{itemize}

\item {\bf Physical attack}~\cite{akhuseyinoglu17, cao20, skandhakumar12}: A physical attack refers to the attack scenario where attackers attempt to break access control systems and physically access CPSs.

\item {\bf Zero-day attack}~\cite{borbor2019optimizing,Carvalho-msthesis14,clark1987comparison, deswarte1998diversity,Garcia2014,Gorbenko2019,huang2014links, liu2008diverse, nagy2006n,reiter1996distributing, reynolds2003line2, reynolds2003line}: This attack utilizes unknown vulnerabilities where patches for the vulnerabilities are not available yet. Diverse but redundant authentications are used to thwart such attacks. The authentication is performed by comparing outputs of diverse implementations of system components~\cite{Carvalho-msthesis14,nagy2006n, reynolds2003line2, reynolds2003line}.
\item {\bf Worm attack}~\cite{bailey2005malware,Cho19,gashi2009experimental,Hole2013,Soodeh16-epidemic,huang2014toward,huang2017software,just2004review,ODonnell05-epidemic,o2004achieving, Calton96, Reynolds2002, temizkan2017software, yang2016improving}: A worm is a malicious computer program that can self-replicate and spread to other network computers. After the worm infects a machine, it can edit a file or monitor the machine.  Software diversity can enhance survivability of the Internet against worm attacks~\cite{bailey2005malware}. Graph coloring algorithms are also used to increase the diversity of software packages assuming that different software packages have the different degree or types of vulnerabilities~\cite{Soodeh16-epidemic,huang2014toward,huang2017software,ODonnell05-epidemic,o2004achieving, temizkan2017software,  yang2016improving}.  Software diversity-based topology adaptation is another effective way to thwart such epidemic attacks~\cite{Cho19}.

\item {\bf Code injection attack}~\cite{Barrantes03, cox2006n, hu2006secure, Kc03, Qu2018,Williams09,Xu03}: This attack injects a payload, which is usually binary, to a running application, and then forces the application to run the payload.  Since the injected code can only work when an environment is compatible, random instructions generated for each program have been proposed when they are loaded to memory~\cite{Kc03}. Since the attacker don't know the randomization value, the attacker cannot execute the payload properly.

\item {\bf Code reuse attack}~\cite{Homescu13, Koo16, Williams09}: This attack changes the function pointer of a program so that the program is going to execute malicious behavior. A profile-guided automated diversity approach to defend against code reuse attacks~\cite{Koo16}.  

\item {\bf Return-to-Libc attack}~\cite{Williams09}:
This attack is often applied when a buffer overflow error occurs. The attacker usually has prior knowledge of the stack address. And then, it replaces the return address with the address of another subroutine, subsequently forcing the application to execute a library function with malicious arguments. To deal with this attack, the calling sequence diversity (CDS) of functions is used~\cite{Williams09}.  Since different programs have their respective calling sequence with CSD, the attack for one program cannot be easily propagated to other programs.  An attacker may guess the key value for each function by constantly observing the function address although the attacker cannot use the same value for other programs.

\item {\bf Correlated attack}~\cite{hong2017optimal}: A correlated attack refers to an attack scenario where one compromised node's failure can cascade to the class of nodes it belongs to immediately. Under this situation, the attacker is assumed to have direct access to all nodes in the network. Graph coloring algorithms are used to optimally assign software variants to the network to maintain the maximum connectivity and security~\cite{hong2017optimal}.

\item {\bf Coordinated attack}~\cite{touhiduzzaman2018diversity}: This type of attack allows attackers to target multiple network assets simultaneously. Thus, software monoculture can introduce significant vulnerability in this scenario. Game theoretic approaches are used to mitigate and thwart this kind of attack by optimally assigning different software packages in the network~\cite{touhiduzzaman2018diversity}.

\item {\bf Buffer overflow attack}~\cite{bittau2014, Chew02, forrest1997building, prasad2003binary,sovarel2005s}: The buffer overflow attack targets services that automatically restart once a machine is crashed.  This attack would first scan and read the stack to identify potential vulnerabilities and then remotely perform the write operation to steal a server's binary code. Note that attackers may overwrite the stack with their guesses until services crash and restart, which allows them to try more without being detected.

\item {\bf Side-channel attack}~\cite{crane2015thwarting, hataba2015diversified, karam17, tromer2010efficient, wang2019diversity, zhou2005}: A side-channel attack targets at the implementation environment of an algorithm other than the algorithm itself.  These environmental factors can be physical, such as power supply and acoustic variables, or non-physical, such as cache and running time during the execution of algorithms.

\item {\bf Deobfuscation attack}~\cite{collberg1998manufacturing, pawlowski2016probfuscation}: This kind of attack tries to perform malicious reverse engineering on obfuscated code.  Specifically, the attacker aims to undo the obfuscating transformations on the original program and retrieve valuable information out of it. The attacker may perform analysis through multiple traces in order to efficiently and effectively deobfuscate the target program~\cite{pawlowski2016probfuscation}.

\item {\bf Impersonation attack}~\cite{cobb2010physical, danev2010attacks, foruhandeh2019simple, gerdes2011physical, gerdes2006device}: This type of attack fools the identification system in order to disguise the malicious behavior, if any. This is normally done by mimicking or replaying the features and signals extracted from normal communications with other nodes~\cite{danev2010attacks, gerdes2011physical}.

\item {\bf Tampering attack}~\cite{karam17}: This attack targets at a system's physical identity information, such as an IP address. For example, the attacker may keep sending requests to the system and analyze the  returning bitstream~\cite{karam17}. After retrieving the identity information, the attacker may tamper the systems' identity and launch other attacks, such as impersonation attack.

\item {\bf Eavesdropping attack}~\cite{ghourab2017towards, ghourab2019spatiotemporal, sarkar2012enhancing, zeng10, zou2015improving}: This attack allows the attacker to passively gain information through the network communications. This can be done by installing malware or injecting a virus into compromised network clients.  The stolen private information would later be stored and analyzed to engage in malicious activities.

\item {\bf Denial of Service attack}~\cite{danev2009transient,gondal2015network, Totel2005}: This type of attack typically sends out an extensive number of requests to servers, which causes a system's temporary overload and dysfunction. Normal valid requests would be rejected during the temporary shutdown period.

\item {\bf Reverse engineering}~\cite{collberg1998manufacturing, Franz10, kuang2016exploiting, xue2018exploiting}: Reverse engineering refers to the process where attackers could analyze and identify system components and their interrelationships so that attackers can further leverage them to reconstruct the system in a similar form.

\end{itemize}

\section{Metrics Used for Diversity-based Approaches}

In this section, as we mentioned in Section VI.A of the main paper, we provide the detail of each metric used in the existing diversity-based approaches.

\subsubsection{\bf Security Metrics}  The existing diversity-based approaches have used the following metrics to measure security:
\begin{itemize}
\item {\bf Epidemic Threshold}~\cite{Soodeh16-epidemic, ODonnell05-epidemic, temizkan2017software}: This metric measures the effectiveness of diversity in terms of the rate of worm or virus propagation in a network~\cite{ODonnell05-epidemic}.  In the context of worm propagation in a network, the rate of propagation is defined as the ratio of the infection rate ($\beta$) to the recovery rate of nodes ($\gamma$). This ratio, $\beta / \gamma$, is compared with the threshold, below which the infection dies out.  An epidemic threshold is considered as:
\begin{equation}
    \frac{\beta}{\gamma} < \frac{N}{\langle k \rangle},
\end{equation}
where $\langle k \rangle$ is the average node degree and $N$ is the number of different software variants available in the network.  This metric well reflects the extent of network security under epidemic attacks.  However, the epidemic threshold is not scalable for large-scale networks, such as scale-free networks~\cite{Soodeh16-epidemic, temizkan2017software}. 

\item {\bf Diversity metrics}: The extent of diversity is measured in code, instructions, or routing paths.  The examples include:
\begin{itemize}
\item {\bf Expected path variant (EPV)}~\cite{hong2017optimal}: This metric measures the expected proportion of attack paths with more than one software variant, indicating the extent of system security.

\item {\bf Randomization code coverage}~\cite{Koo16}: This metric evaluates the percentage of randomized instruction sets in a code base.  This is often used to defend against code reuse attacks where a more randomized code means lower code reusability during attacks.

\item {\bf Code similarity}~\cite{hataba2015diversified,pawlowski2016probfuscation}:
This metric quantifies the effectiveness of code obfuscation techniques by code similarity degree. The similarity degree is represented by the percentage of identical codes in the obfuscated algorithm.
\end{itemize}
\item {\bf Vulnerability metrics}: The metrics to capture system vulnerability (or exploitability) to attacks are:
\begin{itemize}
\item {\bf Percentage of granting requests}~\cite{cao20}: This metric evaluates system vulnerability based on the proportion of granted requests over all access requests in which users with low trust are not granted for their access rights. 

\item {\bf Maximum size of common vulnerability graphs}~\cite{huang2014toward, huang2017software}: This metric measures the degree of system vulnerability in terms of the number of common vulnerability graphs (CVGs) as a subgraph where all nodes share the same software variants and where the size of CVGs is affected by how the software variants are assigned in a network (i.e., a software assignment problem).  \citet{huang2017software} considered three different types of networks where each network is a weighted communication graph. 

\item {\bf Node exploitability index}~\cite{touhiduzzaman2018diversity}: This metric measures a node's degree of being exploitable by its neighboring nodes when they are compromised.  \citet{touhiduzzaman2018diversity} used the sum of normalized exploitability indices of all nodes as a system security metric. 

\item {\bf Common vulnerabilities}~\cite{Garcia2014,Gorbenko2019}: This metric gives the extent of shared common vulnerabilities between pairs of operation systems where the data used by this metric are obtained from the NIST National Vulnerability
Database (NVD).

\item {\bf Connected component count}~\cite{o2004achieving}: This metric gives the total number of connected components in a network, where all nodes in each of these components have the same software package.

\item {\bf Secrecy capacity and intercept probability}~\cite{ghourab2017towards,zou2015improving}:
This metric measures the system confidentiality by the secrecy capacity, which is the difference between the main link and the wiretap link channel capacity. Lower secrecy capacity results in higher intercept probability and thus lower system confidentiality.
\end{itemize}
\item {\bf System compromise metrics}: The extent of compromised nodes or compromised routes in a given system or network are used as a metric:
\begin{itemize}
\item {\bf Fraction of compromised nodes}~\cite{Cho19}: This metric estimates the proportion of compromised nodes under epidemic attacks when a network is adapted based on a software diversity value estimated in terms of the software variants and the software vulnerabilities in each node.

\item {\bf Defective edge count}~\cite{o2004achieving}: Defective edges refer to edges between nodes with the same version of software packages in a given network.  This metric measures the total number of defective edges in the network.
\end{itemize}

\item {\bf Intrusion detection accuracy}~\cite{Barrantes03,cox2006n,danev2009transient,foruhandeh2019simple,gashi2009experimental,gerdes2006device,gondal2015network,gu2008principled,hu2006secure,Kc03,majorczyk2007experiments,nagy2006n, Qu2018,Totel2005}: This metric evaluates the security performance of diversity-based IDS. The higher detection accuracy indicates the contribution of diversity schemes associated with the system~\cite{Qu2018}. Specifically, intrusion detection accuracy can be evaluated by true positive, false positive, true negative, false negative, and equal error rate of IDS.


\end{itemize}

\subsubsection{\bf Dependability Metrics} The following dependability metrics have been used by diversity-based security techniques:
\begin{itemize}
\item {\bf Quality-of-Service (QoS) metrics}: QoS is often measured to indicate system dependability. The examples are:
\begin{itemize}
\item {\bf Total packet loss rate}~\cite{hong2017optimal}: This metric is commonly used to measure service availability of a system, measuring the total number of packets lost over the total number of packets transmitted. \citet{hong2017optimal} used this metric to measure service availability where a network topology is periodically changed in order to change attack paths as in a moving target defense mechanism.

\item {\bf Execution time delay}~\cite{Barrantes03,Homescu13,hu2006secure, Kc03, Koo16, Williams09}: This metric quantifies the average running time of the defense system, which is compared against that of the original system without a defense mechanism. The metric is often represented by the ratio of delay introduced in the proposed defense system over the delay in the original system.

\end{itemize}

\item {\bf Reliability}: Operational system reliability is measured as an aspect of dependability as well.  The examples are as follows:
\begin{itemize}
\item {\bf Loss of load}~\cite{touhiduzzaman2018diversity}: This metric measures how much load has been reduced due to attacks applied to a network of interconnected substations. This metric represents system reliability since lower loss of load indicates high reliability with reduced damage introduced by attacks, which happen after a graph coloring algorithm is applied to adapt an original network to a network with a high network diversity in terms of software variants.

\item {\bf Functional discrepancies}~\cite{liu2008diverse}:
This metric detects and outputs all existing functional discrepancies between multiple firewall versions by a customized comparison algorithm.

\item {\bf Survivability rates}~\cite{Homescu13,Calton96}:
This metric measures system reliability by observing the execution output of some system checkpoints, such as specialized codes~\cite{Calton96} or instruction sequence of gadgets~\cite{Homescu13}. The higher survivability rates of specialized codes indicates higher system reliability against security faults caused by attacks.
\end{itemize}

\item {\bf Maintainability}: Financial cost or overhead is considered to measure dependability as well. The examples include:
\begin{itemize}
\item {\bf Minimum number of software packages}~\cite{temizkan2017software}: This metric measures the minimum number of software packages installed to obtain the administration and support cost along with software diversity maintenance.  To be specific, a system imposes a minimum number of of installations for each software package so that they could control the maintenance cost of software diversity.

\item {\bf Stack usage}~\cite{forrest1997building}:
This metric evaluates the space maintenance cost of diversity mechanisms with additional required stack space needed in the execution process. The additional space is commonly quantified as a percentage of the original stack space.

\end{itemize}
\end{itemize}

\section{Datasets Used for Diversity-based Approaches}

We summarized the datasets used to validate the proposed diversity-based approaches proposed based on 35 research papers in Table~\ref{tab:datasets-2}.

\begin{table*}[t]
\centering
\caption{Characteristics of Datasets Used for Diversity-based Security Approaches \label{tab:datasets-2}} 
\vspace{-2mm}
\begin{tabular}{|P{1cm}|p{6cm}|p{10cm}|}
\hline
{\bf Ref.} & \multicolumn{1}{c|}{\bf Research Goal} & \multicolumn{1}{c|}{\bf Dataset Description} \\
\hline
\hline
\multicolumn{3}{|c|}{\cb {\bf Real-World Datasets}}  \\
\hline
        \cite{lach1999algorithms} (1999) & Evaluate the reliability of proposed approach & Xilinx XC4000 family, Sanders CSRC, and Altera Flex 10k architectures \\
    \hline
        \cite{Barrantes03}  (2003) & Evaluate the Randomized Instruction Set Emulator (RISE) & Public reported attacks (Synthetic machine code injection attacks) for two Linux distributions: RedHat from 6.2 to 7.3 and Mandrake 7.2  \\

\hline
        \cite{Xu03} (2003) & Test the Transparent Runtime Randomization (TRR) &  Publicly vulnerable programs and attack information from security bulletin board in \cite{SecurityFocus} \\
\hline
        \cite{o2004achieving} (2004) & Test the proposed algorithm to enhance network security and limits the ability of malicious node &  Email datasets of the ECE Department at Drexel University, which capture the logs of emails pass though ece.drexel.edu server, which have 1,038,939 logs in total \\

\hline
         \cite{Oberheide2008}  (2008) & Test the proposed antivirus system for in-cloud (CloudAV) & Two datasets: (i) collected malicious software over period of one year; and (ii) collected by scanning a campus network for multiple department over six months \\        
\hline
        \cite{liu2008diverse} (2008) & Evaluate the proposed firewall design method & The firewall is maintained by senior firewall administrator in a university, and the author upgrade the firewall policy with their diverse firewall design method.\\
\hline
        \cite{gashi2009experimental} (2009) & Analyze the potential gains in dataset & SGNET \cite{leita2008sgnet_2} \cite{leita2008sgnet_3} is real world honeypot deployment which contain 1599 malware samples.\\
\hline
        \cite{watteyne2009reliability} (2009) & Replay the behavior of channel hopping MAC protocol in real-world traces & Collected real-world data with 46 IEEE802.15.4-compliant TelosB motes where those motes are deployed in UC Berkeley office space, a 50m $\times$ 50m indoor environment, and constantly listen for pockets and 12 million pockets are collected and used in experiment \\
\hline
        \cite{zeng10} (2010) & Test the proposed Multiple-Antenna KEy generator (MAKE) & Real performance data from three Dell e5400 laptops where the first and the second laptops run MAKE protocol and communicate with 2.4GHz frequency, 12Mbps modulation rate, and 15dBm transmission power while the third laptop eavesdrops the communication \\
\hline
        \cite{Bishop11} (2011) & Analyze the benefit using diversity antivirus product  & 1599 malware from distributed honeypot deployment called SGNET~\cite{leita2008sgnet, canto2008large}, which utilizes a protocol learning technology to observe and emulate a code injection attack \\
 \hline
        \cite{Garcia2014}  (2014) & Analyze OS common vulnerabilities and defense effect of OS combination  & 2120 OS-level vulnerabilities collected from 44,000 vulnerabilities published by NVD (NIST National Vulnerability Database) \\
\hline
        \cite{Carvalho-msthesis14} (2014) & Build real-world scenarios & XML signature wrapping vulnerabilities in Apache Rampart, where the vulnerabilities are found in Amazon Web Server Apache Axis2 framework \\        
\hline
        \cite{crane2015thwarting} (2015) & Evaluate the proposed defense mechanism under side-channel attack & Two side-channel attacks proposed in~\cite{tromer2010efficient}, in which the target of those two attacks in the evaluation is a cryptographic library of GnuPG, which is AES-128 encryption routine in libgcrypt 1.6.1 \\
        
\hline
        \cite{gondal2015network} (2015) & Evaluate IDS performance &  KDDCup99 dataset~\cite{KDDCup99_dataset}, consisting of five categories including Normal, DoS, Probe, User-to-Root, and Remote-to-Local \\
\hline
        \cite{smutz2016tree} (2016) & Evaluate the proposed approach to solve intrusion delection using classifiers &  Two malware detection systems: PDFrate and Drebin where the operational dataset for PDFrate is from a real world sensor and Drebin uses published dataset \\
\hline
        \cite{zhang2016network} (2016) & Test the proposed network diversity metrics & The National vulnerability databases from~\cite{blueonix} on May 9, 2008 \\
\hline
        \cite{kuang2016exploiting} (2016) &  Evaluate the DSVMP (i.e., a novel VM-based code obfuscation) performance & Obtained from a debugging program with IDA~\cite{Ida_software} \\
\hline
        \cite{Qu2018} (2018) & Implement a multi-stage approach for detect injection attacks & MidiCart ASP version with Microsoft SQL server and PHP version with Mysql as datasets where the ASP version and PHP version are obtained from an online shopping cart application, called {\em Midicart}\\        
\hline
        \cite{touhiduzzaman2018diversity} (2018)  & Evaluates the proposed coloring algorithm & IEEE-14 bus and IEEE-118 bus test case system \\
\hline
        \cite{azmoodeh2018SC} (2018) & Test the proposed malware detector for IoBT & 128 malware samples using VirusTotal Threat Intelligence platform, and 1078 goodware from official IoT App store, like Pi Store, where both malware and goodware are designed for ARM-based IoT applications; To obtain the OpCode sequence, Objdump is utilized as a disassembler to extract OpCodes. \\
\hline
         \cite{Gorbenko2019} (2019) & Exam diverse intrusion-tolerant architectures & Vulnerabilities between 1/1/2012 and 12/31/2017 for six OSs (Ubuntu, Red Hat, Novell, Windows, MacOS, and Solaris) \\

\hline
        \cite{alavizadeh2019model} (2019) & Evaluate the deployment for proposed MTD technique & Win10, Linux, and Fedora vulnerabilities from National Vulnerability Database (NVD).\\
\hline
        \cite{wang2019diversity} (2019) & Test MLP and CNN models & Attack traces between malicious and victim devices, collected by a supervised ChipWhisperer board, where those two devices are identical 8-bit Atmel microcontrollers ATxmega128D4 placed on different types of printed circuit boards \\
\hline
\end{tabular}
\vspace{-2mm}
\end{table*}
         
\begin{table*}[t]
\centering
\caption*{\sc (continued) Characteristics of Datasets Used for Diversity-based Security Approaches}
\label{tab:datasets}
\vspace{-2mm}
\begin{tabular}{|P{1cm}|p{6cm}|p{10cm}|}
\hline
{\bf Ref.} & \multicolumn{1}{c|}{\bf Research Goal} & \multicolumn{1}{c|}{\bf Dataset Description} \\
\hline
\hline
\multicolumn{3}{|c|}{\cb {\bf Synthetic Datasets}}  \\
\hline
         \cite{azab2011chameleonsoft} (2011) & Evaluate the proposed defense mechanism called {\em ChameleonSoft} & A synthetic static network with the same parameters for each iteration under shuffling and attack events \\\hline
         \cite{Antunes11} (2011) & Evaluate the proposed methodology based on the compliance of diverse server replicas &  Synthetic network traffic data from scripted clients which build with different protocols such as FTP, SMTP, or POP \\
         \hline
        \cite{huang2014toward} (2014) & Evaluate the performance of a set of off-the-shelf software selected based on the proposed algorithm  & Three synthetic graphs with different topologies, including random, regular, and power-law graphs. 
        \\
\hline
         \cite{temizkan2017software}  (2017) & Validate the proposed optimization model for software distribution &  Synthetic network dataset generated by Network Workbench~\cite{nwb2006network} with size 1,000, 2,000, and 5,000 nodes based on scale-free network model~\cite{barabasi1999emergence}.\\
\hline
         \cite{huang2017software}  (2017) & Evaluate the performance of the proposed software assigning algorithm & Three different graph models used to generate three types of synthetic graphs (i.e., random graphs, scale-free graphs, and small world graphs) \\ 
\hline
         \cite{hong2017optimal} (2017) & Evaluate the proposed network topology shuffling algorithm in an SDN testbed & Generated 1000 random topologies with different node and network density \\
\hline
         \cite{Cho19} (2019) & Evaluate the proposed software diversity adaptation schemes & Synthetic network datasets for different network topologies, such as a random network and a scale-free network \\
    \hline
    \hline

\multicolumn{3}{|c|}{\cb {\bf Semi-Synthetic Datasets}}  \\
        \hline
         \cite{Reynolds2002} (2002) & Evaluate the proposed defense approach under Code Red I and II attacks & Datasets from the HACQIT web server where Code Red I, Code Red II attacks are collected from from both an ordinary laptop and the log generated by the laptop \\
                \hline
\cite{hu2006secure} (2006) & Evaluate the proposed approach under code-injection attacks & Published vulnerabilities used as seeds, which are combined with synthetic vulnerabilities into real applications \\
        \hline
         \cite{gu2008principled}  (2008) & Evaluate the proposed fusion technique based on LRT (likelihood ratio test) & Two datasets: (i) Network traffic dataset generated from DARPA Intrusion Detection Evaluation program in 1998, which is often called the KDD Dataset; and (ii) Real traces collected from their campus web server based on the observations of 30 minutes HTTP traffic, which contains 5 million packets \\
         \hline
         \cite{Homescu13}  (2013) & Evaluate the proposed profile-guided optimization &  Synthetic datasets by collecting profile information from SPEC (Standard Performance Evaluation Corporation) CPU 2006 with trained input set \\
\hline
        \cite{borbor2019optimizing}  (2019) & Evaluate the proposed automated approach to improve network resilience under unknown attack & Synthetic graphs by choosing realistic networks as seed graphs where nodes and edges are added randomly \\
        
\hline
\end{tabular}
\vspace{-2mm}
\end{table*}

\section{Validation Testbeds Used for Diversity-based Approaches}

In this section, we survey the evaluation testbeds used to validate the existing diversity-based approaches for secure and dependable CPSs. We discuss the four types of evaluation testbeds used in the literature: analytical model-based, simulation-based, emulation-based, and real testbed-based.

\subsubsection{\bf Analytical Model-based Validation}
A stochastic model has been used to model a system, attack behaviors, and interactions of system components in order to assess the proposed technique. These models provide mathematical proof of concept for the proposed technique.  \citet{sarkar2012enhancing} studied the effect of antenna diversity by analytically modeling secrecy capacity based on probabilistic methods. \citet{zou2015improving} evaluated the secrecy capacity and the intercept probability for different number of relays mathematically.  \citet{Gorbenko2019} statistically analyzed multiple Operating Systems to study the impact of $N$-version OS diversity in intrusion detection.

\subsubsection{\bf Simulation-based Validation}
In simulation models, the real-world systems, attackers, and their interactions are imitated using software~\cite{ghourab2017towards, ghourab2019spatiotemporal, karam17}. These models provide more flexibility in evaluating the system under various attack scenarios and related parameters.  Among the hardware diversity techniques, \citet{karam17} used a FPGA mapping tool to evaluate their proposed architectural diversity technique against brute force and side-channel attacks. \cite{Calton96} also developed a simulation model for the diversified implementation of OS kernels against simulated attack scenarios.  Some network diversity approaches used network analysis tool-kits for simulating computer networks as graphs, such as the Network Workbench tool~\cite{touhiduzzaman2018diversity} or the virus propagation on a diversified network~\cite{nwb2006network}.

\subsubsection{\bf Emulation-based Validation} While the simulation-based models provide an approximation of the real-world systems, emulation-based models more accurately replicate the real-world systems in a virtual environment.  These models provide stronger validation of the results. Although emulation and virtualization have few differences in their definitions within the scope of this paper, both the methods are included under emulation-based validation. Therefore, the techniques evaluated using virtual machine implementation are included here.  \citet{Kc03} implemented their proposed instruction-set randomization based approach on an x86 emulator provided by \textit{bochs}~\cite{bochs}.  \citet{Silva2010} prototyped a replication-based anti-virus engine on {\em Xen}~\cite{xen} virtualization software.

\subsubsection{\bf Real Testbed-based Validation} PLI techniques for vehicular networks have been tested on TM4C123GXL micro-controllers~\cite{foruhandeh2019simple, lach1999algorithms, wang2019diversity, zeng10}.  \citet{gerdes2006device} performed the validation for their proposed PLI technique for Ethernet devices on two CPUs and measurement setup.  To study the effect of frequency hopping on the network routing,  \citet{watteyne2009reliability} analyzed the network traces from a real-world deployment of IEEE 802.15.4-compliant motes in an indoor environment. They also studied the interference of IEEE 802.11 (WiFi) signals in their experiments.  \citet{gashi2009experimental} evaluated the different antivirus engines against various malware on a web service called VirusTotal~\cite{virustotal}. \citet{liu2008diverse} tested the effectiveness of diverse Firewall design using both real and synthetic Firewalls. \citet{Reynolds2002} analyzed the effectiveness of their proposed intrusion detection system by launching attacks from a malicious laptop.

\bibliographystyle{IEEEtranSN}
\bibliography{ref}